\documentclass[prd,twocolumn,floatfix,nofootinbib]{revtex4}
\usepackage{amssymb}
\usepackage{amsmath}
\usepackage{graphicx,subfigure,color,dcolumn,booktabs,bm}
\usepackage{longtable,lscape}
\usepackage{txfonts}
\usepackage{overpic}
\usepackage{indentfirst}
\usepackage{cases}
\usepackage{multirow}
\usepackage{ulem}
\usepackage[colorlinks,
            citecolor=blue,
            anchorcolor=red,
            menucolor=red,
            linkcolor=red,
            filecolor=red,
            runcolor=red,
            urlcolor=blue,
            frenchlinks=false]{hyperref}

\graphicspath{{Figures/}}
\allowdisplaybreaks
\begin{document}

\title{Prospects for compact hexaquarks under the limitation imposed by quark confinement}
\author{Wen-Xuan Zhang$^{1,2,3}$}
\email{zhangwx89@outlook.com}
\author{Wen-Nian Liu$^{4,5}$}
\email{1169994277@qq.com}
\author{Duojie Jia$^{2,6}$}
\email{djjia@qhit.edu.cn}
\affiliation{$^1$School of Physical Science and Technology, Lanzhou University, Lanzhou 730000, China \\
$^2$Lanzhou Center for Theoretical Physics, Key Laboratory of Theoretical Physics of Gansu Province, \\
Key Laboratory of Quantum Theory and Applications of MoE, \\
Gansu Provincial Research Center for Basic Disciplines of Quantum Physics, Lanzhou University, Lanzhou 730000, China \\
$^3$Research Center for Hadron and CSR Physics, Lanzhou University and Institute of Modern Physics of CAS, Lanzhou 730000, China \\
$^4$Institute of Theoretical Physics, College of Physics and Electronic Engineering, 
Northwest Normal University, Lanzhou 730070, China \\
$^5$Xinjiang Laboratory Phase Transitions and Microstructures in Condensed Matters, College of Physical Science and Technology, 
Yili Normal University, Yining 835000, China\\
$^6$General Education Center, Qinghai Institute of Technology, Xining 810000, China}
\date{\today}

\begin{abstract}
The limitation of flavor constituents for compact multiquarks is crucial for understanding the strong interaction at the low energy scale.
Utilizing the MIT bag model that incorporates perturbative interactions and confinement energy $E_{\rm CON}$, 
we derive a critical bag radius $R_c=5.61\,$GeV$^{-1}$ from the condition $E_{\rm CON} < 0$ at zero temperature and zero baryon density, 
which aligns with the string-breaking distance of 1.2--1.4$\,$fm. 
Applying this framework to 6-, 7-, and 8-quark systems, we find the bag radii $R_0$ to be highly sensitive to relativistic effects from light quarks, 
leading to the exclusion of most heavy-light flavor configurations (e.g., $n^3\bar{c}^3$, $n^3\bar{n}\bar{c}^2$) due to positive $E_{\rm CON}$ and radii exceeding the critical radius. 
Color-spin wavefunctions are constructed using Young tableaux to evaluate interaction matrices and OZI-superallowed decays.
Broad decay widths in fully heavy systems for OZI-superallowed modes could arise from wavefunction overlaps due to heavy flavor symmetry, 
suggesting possible narrow widths for $nnb\bar{b}\bar{b}\bar{b}$ and $nnn\bar{b}\bar{b}\bar{b}$ hexaquarks. 
This phenomenological approach provides insights into the limitations on multiquarks imposed by confinement. It recommends experimental searches at LHCb for these states.
\end{abstract}

\maketitle
\date{\today}

\section{Introduction}
\label{sec:intro}

Since the proposal of quark model by Gell-Mann and Zweig in 1964 \cite{gell-mann:1964ewy,zweig:1964jf}, 
the study of multiquark states such as tetraquarks $q^2\bar{q}^2$ and pentaquarks $q^4\bar{q}$, beyond conventional mesons $q\bar{q}$ and baryons $qqq$, 
has received great attention in hadron physics \cite{Liu:2024uxn,Chen:2022asf,Guo:2017jvc,Liu:2019zoy,Chen:2016qju,Hosaka:2016pey,Jaffe:2004ph}.
Over the past two decades, collaborations such as LHCb, Belle and BESIII have reported findings of multiquark states,
including the heavy tetraquarks \cite{Belle:2003nnu,BESIII:2013ris,Belle:2013yex,lhcb:2021vvq,lhcb:2020bwg,CMS:2023owd} and $P_{c}/P_{cs}$ pentaquarks \cite{lhcb:2015c,lhcb:2019kea,lhcb:2020jpq,lhcb:2022ogu}.
In addition to these advances, quark confinement imposes critical scale for strong interactions between quarks.
It becomes timely and crucial to question if there is a size upper limit for multiquark states to spatially contain as many valence quarks as possible. 
The investigation of compact hexaquarks establishes a robust framework for this discussion.

Theoretical progress in hexaquark studies has illustrated their potential structures across compact and molecular scenarios.
With compact interpretations, recent computations employed the chromomagnetic interaction (CMI) to predict mass spectra and relative decay widths for hexaquarks \cite{Liu:2021gva,Liu:2022rzu,Weng:2022ohh,Weng:2024qly,An:2025rjv}.
Light hexaquarks were explored in a QCD-motivated van der Waals model \cite{Andrew:2023aes} and using QCD sum rules \cite{Zhang:2025vqg}.
Within molecular scenarios, the structures of dibaryon \cite{Li:2012bt,Huang:2013rla,Meng:2017udf,Meng:2019nzy,Wang:2019gal,Clement:2020mab,Chen:2021cfl,Chen:2022iil,Bashkanov:2023yca,Zhang:2025ame},
baryon-antibaryon \cite{Lee:2011rka,Lu:2017dvm,Yang:2018amd,Cheng:2022vgy,Kong:2022rvd,Kong:2023dwz,Wang:2024riu,Wu:2024trh}, 
and trimeson \cite{Debastiani:2017vhv,Huang:2019qmw,Wu:2020job,Wu:2021kbu,Luo:2021ggs,Wei:2022jgc,Bayar:2022bnc,Ikeno:2022jbb} were predicted.
Bound and resonant states are discussed for doubly heavy tetraquarks in terms of the size estimations of the substructures to identify compact, molecular and resonant states,
with the scale $\Lambda^{-1}_{\rm{QCD}}\sim 1\,$fm \cite{Wu:2024zbx}. 
Scale of confinement is computed by using lattice QCD \cite{bali:2000gf,Bali:1998de,bali:2005fu,bulava:2019iut}, AdS/QCD \cite{Jiang:2023lmj}, 
and others \cite{Castorina:2007eb,Bonati:2020orj,Kou:2024dml} based on the breaking of string or flux-tube, and suggested to have critical size about 1.2--1.4$\,$fm for inter-quark separation.

Multiquark states were also explored using the MIT bag model in early 1970's \cite{jaffe:1976ig,Jaffe:1976yi,Squires:1979,strottman:1979qu}. 
The same model was later applied to study doubly heavy baryons \cite{fleck:1989mb,he:2004px,bernotas:2008bu,bernotas:2008fv,bernotas:2012nz}, hybrid mesons \cite{barnes:1977hg,barnes:1982tx,chanowitz:1982qj}, 
multiquark spectra and magnetic moments \cite{zhang:2021yul,zhu:2023lbx,yan:2023lvm,zhang:2023hmg,zhang:2023teh,Liu:2024mwn,Liu:2025fbe}, and the semileptonic decays \cite{Geng:2020ofy,Geng:2022uyy,Liu:2022pdk}, among others.   
Knowledge of bag constant $B$ arising from vacuum condensations \cite{Novikov:1981xi,Li:1990es,Buballa:2003qv,Mo:2010zza,Baym:2017whm} and Casimir energy in this model, 
emerges to be useful to understand quark confinement and deconfinement picture, where the volume energy ($\sim B$) and Casimir energy form a negative confinement energy of hadron bag.
The similar issue is also explored in the picture of confinement with domain walls \cite{Mandelstam:1974pi,tHooft:1981bkw,Wang:2007ng,Sulejmanpasic:2016uwq,ma:2019xtx}, 
where surface of hadron bag is superconductive \cite{Skagerstam:1979ka,Nair:1984cf,Ball:1987cf,Ansoldi:1995by,Asorey:2013wvh}.

Moreover, as a function of bag constant $B$, the bag radius $R$ at the surface varies upon temperature and baryon density,
approaching infinity at the critical temperature $T_c$ as $B\to 0$, corresponding to deconfinement.
The temperature dependence of $B$ is detailed in Ref.\cite{Mo:2010zza}, where the bag constant vanishes and the radius becomes infinite at $T_c=133\,$MeV.
For other studies, see Refs.\cite{Muller:1980kf,Plumari:2011mk,Klahn:2016uce,Wen:2024bfv}.
The density dependence indicates an exponentially decreasing behavior of $B$ with baryon density \cite{Liu:2001em,Prasad:2003bw,Pal:2023dlv,Sedaghat:2024rjk,Ju:2025mig}, 
in study of quark stars. In particular, Ref.\cite{Liu:2001em} proposes a critical bag radius $R_c$ in nuclear matter relative to nuclear density.
Upon the advances of limitation research, we implement the critical bag radius $R_c$ at zero temperature and zero baryon density for investigating single hadron properties.

The purpose of this work is to explore stability of compact hexaquarks and more complex multiquark states in consideration of a confinement scale within the MIT bag model. 
Utilizing the mass equation that contains effective confinement energy, we propose a critical bag radius $R_c$ with respect to bag constant 
and apply it to study stability of 6-, 7-, and 8-quark systems of hadrons, including fully heavy and partially heavy systems.
To address gluon exchange interactions and OZI-superallowed strong decays, we construct color-spin wave functions for hexaquarks using Young tableaux, 
expressed in terms of baryon-antibaryon coupling bases, to evaluate interaction matrices and wavefunction overlaps. The stability under confinement and OZI-superallowed decays will 
be related to the effects of light degrees of freedom.

This work is structured as follows. 
In Sec.~\ref{sec:model}, we describe the mass equation of MIT bag model incorporating perturbative interactions and confinement energy. 
Sec.~\ref{sec:confine} discusses the confinement in terms of the bag radius, including the physical picture of the bag model and derivation of the critical radius $R_c$. 
The value of $R_c$ is applied to estimate the stability of multiquark states.
In Sec.~\ref{sec:hexaquark}, we investigate hexaquarks with OZI-superallowed strong decays, focusing on fully heavy systems in subsection A and partially heavy systems in subsection B. 
Finally, a summary and conclusions for search of hexaquarks are presented in Sec.~\ref{sec:summary}.

\section{The method of MIT bag model}
\label{sec:model}

Considering heavy hexaquark system as a spherical bag with radius $R$ in the framework of MIT bag model, which implements the notions of confinement and asymptotic freedom simultaneously,
the mass of bag system in MIT bag model is given by
\begin{equation}
    M = \sum_i\omega_i+E_{\rm{CON}}+E_{\rm{CMI}}+E_{\rm{BD}},
    \label{equ:mass}
\end{equation}
separated into terms representing various physical effects, for the better understanding of hadron spectra.

The leading-order term of Eq.(\ref{equ:mass}), $\sum_i \omega_i$ stands for the sum of kinetic energy of quarks, where $\omega_i=(m_i^2+x_i^2/R^2)^{1/2}$ indicates the quark $i$ moving relativistically
with mass $m_i$ and momentum $x_i/R$. The values of the quark mass $m_i$
were determined in our previous work \cite{zhang:2021yul} using light baryons and heavy mesons \cite{particledatagroup:2022pth}. 
The dimensionless variable $x_i$ is solved with the boundary condition of quark spinor wavefunction $\psi_{i}(r)$ at the bag surface $r=R$ \cite{greineri2007}:
\begin{equation}
    \begin{aligned}
        &\left.
        \begin{pmatrix}
            0 & -i\sigma_r \\
            i\sigma_r & 0
        \end{pmatrix} \psi_{i}(r) = \psi_{i}(r) \right|_{r=R} \\[2mm]
        &\implies \tan x_{i}=\frac{x_{i}}{1-m_{i}R-\left( m_{i}^{2}R^{2}+x_{i}^{2}\right)^{1/2}},
    \end{aligned}
    \label{equ:boundary}
\end{equation}
where 
\begin{equation}
    \psi_{i}(r) = N_{i} \binom{j_{0}(p_{i}r)\chi^{\mu}_{\kappa}}
    {i\frac{p_{i}}{\omega_{i}+m_{i}}j_{1}(p_{i}r)\boldsymbol{\sigma}\cdot\boldsymbol{\hat{r}}\chi^{\mu}_{\kappa}} e^{-i\omega_{i}t},
    \label{equ:quarkspinor}
\end{equation}
with $j_l$ the Bessel function and $p_i=x_i/R$ the quark momentum.
Numerically, $x_i$ is fixed at 2.04279 in the chiral limit $m_u=m_d=0\,$MeV, and approaches $\pi$ in heavy quark limit.

The second term of Eq.(\ref{equ:mass}) is the bag confinement energy $E_{\rm{CON}}$ arising from QCD vacuum effects. This term has two contributions:
\begin{equation}
    E_{\rm{CON}} = \frac{4}{3}\pi R^{3}B-\frac{Z_{0}}{R}, 
    \label{equ:bconf}
\end{equation}
where the first is volume energy with bag constant $B$ derived from quadratic boundary condition, 
and the second is Casimir zero point energy characterized by the free parameter $Z_0$. The values of both parameters can be found in Ref.\cite{zhang:2021yul},
where a good overall fit of light and heavy hadron spectra is obtained.
This part of energy serves to provide long-range binding effect and form the bag to be sphere and dynamically stable.
Generally, $E_{\rm{CON}}$ should be negative to guarantee effective binding energy and thus enforce quark confinement.
However, in other cases, they could be largely positive, indicating challenges in hadronization.
We discuss this further in the next section.

The chromomagnetic interaction (CMI), $E_{\rm{CMI}}$, stemming from the lowest-order gluon exchanges \cite{derujula:1975qlm,Liu:2019zoy}, 
takes a general form
\begin{equation}
    E_{\rm{CMI}}=-\sum_{i<j}\left\langle \boldsymbol{\lambda _{i}}\cdot \boldsymbol{\lambda
    _{j}}\right\rangle \left\langle \boldsymbol{\sigma _{i}}\cdot \boldsymbol{\sigma _{j}}
    \right\rangle C_{ij}, 
    \label{equ:CMI}
\end{equation}
where $i$ and $j$ denote quarks or antiquarks, $\lambda$ represents the Gell-Mann matrices, $\sigma$ stands for the Pauli matrices,
and $C_{ij}$ refers to the CMI parameters. In the case of MIT bag model, this parameter is given by \cite{degrand:1975cf}
\begin{equation}
    C_{ij}=3\frac{\alpha _{s}\left( R\right) }{R^{3}}\bar{\mu}_{i}\bar{\mu}_{j}I_{ij},
    \label{equ:Cij}
\end{equation}
with running strong coupling $\alpha _{s}(R)$ and quark magnetic moment $\bar{\mu}_{i}$ given by Refs.\cite{zhang:2021yul,yan:2023lvm,zhang:2023hmg}.
The expectation values of Casimir operators in Eq.(\ref{equ:CMI}) referred to as color and spin factors, can be computed using the following relations
\begin{equation}
    {\left\langle \boldsymbol{\lambda }_{i}\cdot \boldsymbol{\lambda }_{j}\right\rangle }_{nm}
    = \sum_{\alpha =1}^{8}\mathrm{Tr}\left( c_{in}^{\dagger }\lambda^{\alpha }c_{im}\right) \mathrm{Tr}\left( c_{jn}^{\dagger }\lambda ^{\alpha}c_{jm}\right), 
    \label{equ:colorfc}
\end{equation}
\begin{equation}
    {\left\langle \boldsymbol{\sigma }_{i}\cdot \boldsymbol{\sigma }_{j}\right\rangle }_{xy}
    = \sum_{\alpha =1}^{3}\mathrm{Tr}\left( \chi _{ix}^{\dagger }\sigma^{\alpha }\chi _{iy}\right) \mathrm{Tr}\left( \chi _{jx}^{\dagger }\sigma^{\alpha }\chi _{jy}\right),
    \label{equ:spinfc}
\end{equation}
where $n$, $m$ and $x$, $y$ describe the color-spin subspace for hadronic state,
and the symbols $c$ and $\chi$ denote color and spin vectors associated to a quark $i$ or $j$.
With the help of color-spin wavefunctions given in Appendix B, one can evaluate the matrix elements of Eq.(\ref{equ:CMI}) explicitly and further diagonalize the mass matrix (\ref{equ:mass}).

The fourth term $E_{\rm{BD}}$ is an enhanced binding energy existing between heavy-heavy or heavy-strange quarks, proposed in Refs.\cite{karliner:2014gca,karliner:2017elp}.
Arising from short-range gluon exchanges, it describes additional strong binding between two quarks of heavy flavor or of one heavy and one strange.
We introduced and extracted five binding energies $B_{cs}$, $B_{cc}$, $B_{bs}$, $B_{bc}$ and $B_{bb}$ for color $\bar{3}_c$ representation in Ref.\cite{zhang:2021yul} based on measured data.
For general color rep. $\mathbf{R}$, we apply the feature of short distance force in QCD and utilize the color factor as scaling to obtain the binding energy \cite{karliner:2014gca,zhang:2021yul}
\begin{equation}
    E_{\rm{BD}} 
    = \sum_{i<j} B_{[ij]\mathbf{R}}
    = \sum_{i<j} \frac{{\left\langle\boldsymbol{\lambda_{i}\cdot\lambda_{j}}\right\rangle}_{\mathbf{R}}}
        {{\left\langle\boldsymbol{\lambda_{i}\cdot\lambda_{j}}\right\rangle}_{\mathbf{\bar{3}}}}B_{ij},
    \label{equ:bindings} 
\end{equation}
where $B_{ij}$ are the binding energies for quark pair $ij$ in $\bar{3}_c$ rep. in the baryons. 
Similarly, binding energies in Eq.(\ref{equ:mass}) may take matrix form in the given color-spin subspace.

\section{Confinement upon bag radius}
\label{sec:confine}

We now have three major physical effects in this model: quark kinetic $\sum_i \omega_i$, perturbative gluon exchanges $E_{\rm{CMI}}+E_{\rm{BD}}$, 
and vacuum energy at the bag level $E_{\rm{CON}}$. The first two are familiar and play a significant role in study of hadron spectrum, involving methods such as
CMI model \cite{derujula:1975qlm,Liu:2019zoy}, potential model \cite{Eichten:1974af,Eichten:1978tg}, and QCD string picture \cite{Nambu:1974zg}.
However, the understanding of non-perturbative QCD and quark confinement remains incomplete. 
The MIT bag model incorporates elements of confinement through the bag radius $R$ and vacuum energy $E_{\rm{CON}}$.
In this section, we discuss the confinement in terms of the bag radius.

Compared to the potential model \cite{Eichten:1974af,Eichten:1978tg} and QCD sum rules \cite{reinders:1984sr,colangelo:2000dp}, 
where the quarks and gluons are confined by long-range strong interactions or vacuum condensations,
the MIT bag model features a distinct hadron surface at the boundary of confinement. At the surface $r=R$,
the vector current $\bar{\psi}\gamma_{\mu}\psi$ satisfies the boundary condition (\ref{equ:boundary}) in the normal direction \cite{jaffe:1976ig,degrand:1975cf}. 
Beyond this mathematical representation of confinement, 
there is a quadratic boundary condition that clarifies the picture.
The surface forms between two phases: non-perturbative QCD filled with vacuum condensations and perturbative QCD occupied by Fock states.
Therefore, the difference in vacuum energy density (bag constant $B$) balances internal bag pressure, ensuring dynamic stability.

Notably, the bag radius $R$ is a variational parameter determined by minimizing the mass equation (\ref{equ:mass}).
Given the quadratic boundary condition, the bag radius is solved variationally so that the bag pressure from quarks inside balance the vacuum energy density outside.
and acquires a stable value $R_0$ as hadron size and energy scale, allowing computation of the strong coupling $\alpha_s(R)$ and each term in Eq.(\ref{equ:mass}).
Every hadron bag is characterized by a unique scale $R_0$, resulting in the mean field $\alpha_s(R_0)$, the mean quark separation $R_0$ as shown in Eq.(\ref{equ:Cij}),
and the exact vacuum energy $E_{\rm{CON}}(R_0)$.

\begin{figure}[t]
    \centering
    \includegraphics[width=0.45\textwidth]{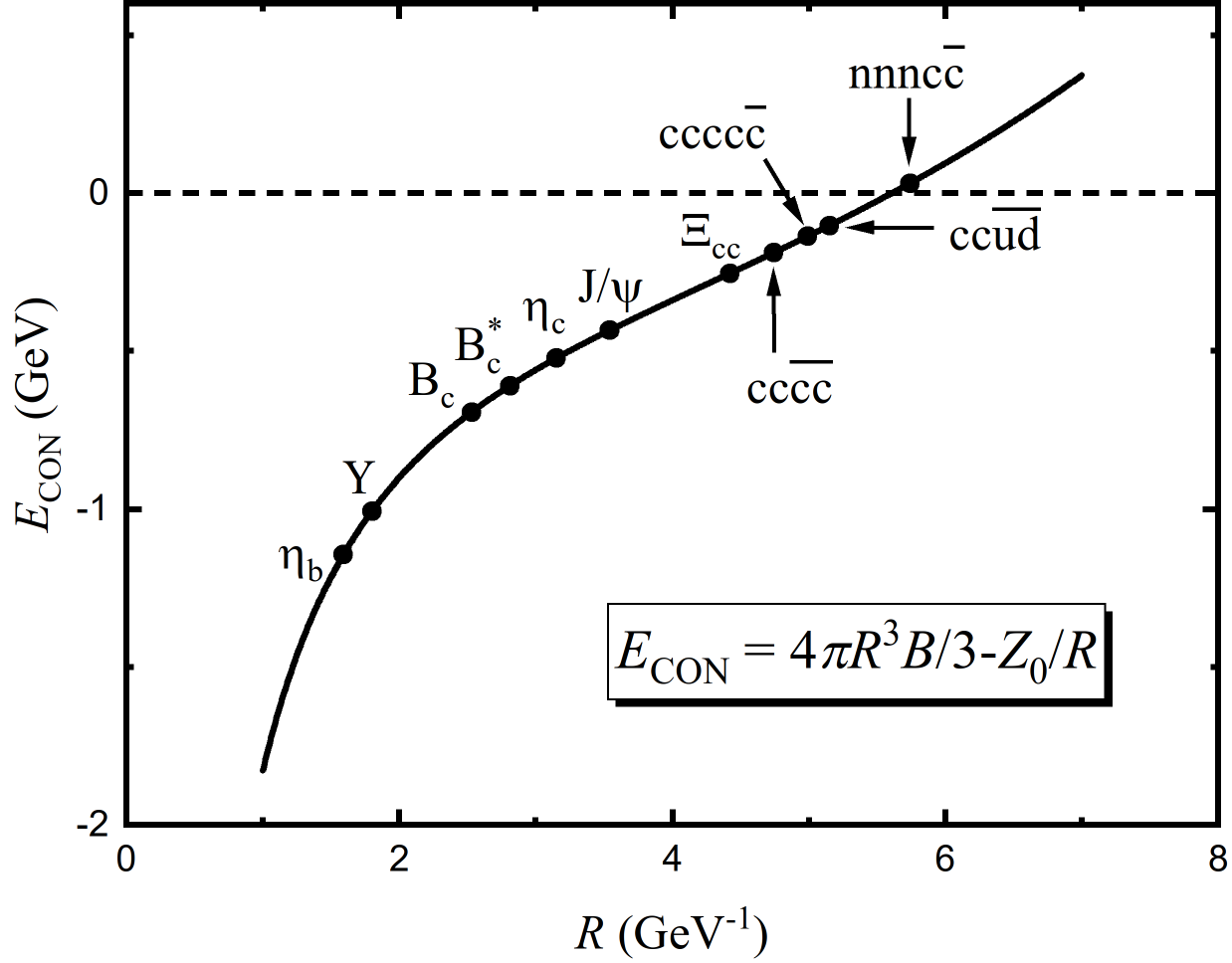}
    \caption{The curve of bag confinement energy $E_{\rm{CON}}$ (\ref{equ:bconf}) as a function of bag radius $R$, with particular hadronic states labeled as sphere dots,
        including heavy quarkoniums, doubly heavy baryon, tetraquarks and pentaquarks.}
    \label{fig:bconf}
\end{figure}

Thus far, we have explained the connection from $E_{\rm{CON}}$ to the bag radius, the quadratic boundary condition, and non-perturbative QCD.
The volume energy in Eq.(\ref{equ:bconf}) tends to expand the bag surface during variation, increasing the mass spectrum, while the zero point energy maintains the bag's integrity.
Together, they play a crucial role in confinement by functioning as binding energy.
Therefore, the confinement energy $E_{\rm{CON}}$ should be negative to form a compact state, via
\begin{equation}
    \begin{aligned}
        E_{\rm{CON}} &= \frac{4}{3}\pi R^{3}B-\frac{Z_{0}}{R} < 0 \\
        &\Rightarrow R < R_c = {\left(\frac{3Z_0}{4\pi B}\right)}^{1/4} = 5.61\mathrm{\,GeV}^{-1},
    \end{aligned}
    \label{equ:critical}
\end{equation}
with $R_c=5.61\,\rm{GeV}^{-1}(1.11\,\rm{fm})$ as the critical radius. We plot the confinement energy $E_{\rm{CON}}$ (\ref{equ:bconf}) in Fig.(\ref{fig:bconf}) with some significant states labeled.
Almost all ground-state hadrons fall within the $R_c$ according to MIT bag model, such as heavy mesons and baryons, doubly heavy tetraquarks \cite{zhang:2021yul}, 
triply heavy tetraquarks \cite{zhu:2023lbx} and pentaquarks \cite{Liu:2024mwn,Liu:2025fbe}, fully heavy tetraquarks \cite{yan:2023lvm} and pentaquarks \cite{zhang:2023hmg}, 
with slight violation for $P_{c}$ states, which are considered unlikely to be compact \cite{zhang:2023teh}. 
The critical radius $R_c$ supports the pictures and conclusions from our previous works.

It is not unique, but the non-perturbative QCD requires the confinement scale of about 1$\,$fm (200$\,$MeV).
For example, lattice QCD predicted a critical string distance to be $r_c\approx 1.2\,$fm \cite{bali:2000gf,Bali:1998de}, 
close to the critical bag radius $R_c$ in this work. Other works for critical distance ranging from 1.2 to 1.4$\,$fm are listed collectively in Table \ref{tab:comparerc}.
It is of interest to compare the bag picture with the string (or flux tube) picture supported by lattice QCD, that the bag can deform and dissociate into separated bags by creating new $q\bar{q}$ pairs \cite{barnes1983241}, 
similar to string-breaking situation. The string picture describes two-body strong interactions for each quark pair, while the bag model absorbs the long-range contributions into vacuum energy at the bag level, 
both reflecting the non-perturbative effects. Therefore, we regard the critical bag radius $R_c$ to be consistent with the critical string distance $r_c$, with minor discrepancies from the spherical cavity and mean field ansatz \cite{degrand:1975cf}.

\renewcommand{\tabcolsep}{0.4cm} \renewcommand{\arraystretch}{1.2}
\begin{table}[!htbp]
    \caption{Comparisons of critical string-breaking distance $r_c$ from various works.}
    \label{tab:comparerc}
    \begin{tabular}{lccc}
    \hline\hline
        Work & Year & $r_c$ & Method \\ \hline
        \cite{Bali:1998de} & 1998 & $\approx 1.2\,$fm & Lattice QCD \\
        \cite{bali:2005fu} & 2005 & $1.248(13)\,$fm & Lattice QCD \\
        \cite{Castorina:2007eb} & 2007 & $\approx 1.2\,$fm & Hawking radiation \\
        \cite{bulava:2019iut} & 2019 & $1.224(15)\,$fm & Lattice QCD \\
        \cite{Bonati:2020orj} & 2020 & $\approx 1.26\,$fm & Yang-Mills theory \\
        \cite{Kou:2024dml} & 2024 & $1.294(40)\,$fm & Schwinger model \\
    \hline\hline
    \end{tabular}
\end{table}

\renewcommand{\tabcolsep}{0.25cm} \renewcommand{\arraystretch}{1.2}
\begin{table*}[!htbp]
    \centering
    \caption{Possible configurations of multiquark states with baryon number $B=0,1$. $R_0$ is the bag radius in unit of GeV$^{-1}$.
        The symbol $\ast$ indicates that the corresponding state is not recommended for study due to the large hadron radius.
        For those with radius larger than 6$\,$GeV$^{-1}(\approx 1.2\,\rm{fm})$, they could be forbidden by quark confinement and are labeled with $\times$.}
    \label{tab:prospect}
    \begin{tabular}{lllllllllllll}
    \hline\hline
        & $n$-$c$ & $R_0$ & $n$-$b$ & $R_0$ & $s$-$c$ & $R_0$ & $s$-$b$ & $R_0$ & $c$-$b$ & $R_0$ & $n$-$c$-$b$ & $R_0$ \\ \hline
        8-quark
        & $c^4\bar{c}^4$ & 5.61$^{\ast}$ & $b^4\bar{b}^4$ & 4.17 & & & & & & & $nc^3\bar{b}^4$ & 5.33 \\
        & $nc^3\bar{c}^4$ & 5.83$^{\ast}$ & $nb^3\bar{b}^4$ & 4.79 & $sc^3\bar{c}^4$ & 5.85$^{\ast}$ & $sb^3\bar{b}^4$ & 4.83 & $cb^3\bar{b}^4$ & 4.44 & $c^4\bar{n}\bar{b}^3$ & 5.47 \\
        & $n^2c^2\bar{c}^4$ & 6.05$^{\times}$ & $n^2b^2\bar{b}^4$ & 5.29 & $s^2c^2\bar{c}^4$ & 6.07$^{\times}$ & $s^2b^2\bar{b}^4$ & 5.35 & $c^2b^2\bar{b}^4$ & 4.67 & $n^2c^2\bar{b}^4$ & 5.59 \\
        & $n^3c\bar{c}^4$ & 6.25$^{\times}$ & $n^3b\bar{b}^4$ & 5.72$^{\ast}$ & $s^3c\bar{c}^4$ & 6.28$^{\times}$ & $s^3b\bar{b}^4$ & 5.78$^{\ast}$ & $c^3b\bar{b}^4$ & 4.87 & $nc^3\bar{n}\bar{b}^3$ & 5.72$^{\ast}$ \\
        & $n^4\bar{c}^4$ & 6.45$^{\times}$ & $n^4\bar{b}^4$ & 6.08$^{\times}$ & $s^4\bar{c}^4$ & 6.48$^{\times}$ & $s^4\bar{b}^4$ & 6.13$^{\times}$ & $c^4\bar{b}^4$ & 5.05 & $c^4\bar{n}^2\bar{b}^2$ & 5.84$^{\ast}$ \\
        & $n^4\bar{n}\bar{c}^3$ & 6.64$^{\times}$ & $n^4\bar{n}\bar{b}^3$ & 6.39$^{\times}$ & $s^4\bar{s}\bar{c}^3$ & 6.67$^{\times}$ & $s^4\bar{s}\bar{b}^3$ & 6.44$^{\times}$ & $c^4\bar{c}\bar{b}^3$ & 5.21 & $n^3c\bar{b}^4$ & 5.84$^{\ast}$ \\
        & $n^4\bar{n}^2\bar{c}^2$ & 6.82$^{\times}$ & $n^4\bar{n}^2\bar{b}^2$ & 6.67$^{\times}$ & $s^4\bar{s}^2\bar{c}^2$ & 6.84$^{\times}$ & $s^4\bar{s}^2\bar{b}^2$ & 6.71$^{\times}$ & $c^4\bar{c}^2\bar{b}^2$ & 5.35 & $c^4\bar{n}^3\bar{b}$ & 6.16$^{\times}$ \\
        & $n^4\bar{n}^3\bar{c}$ & 6.99$^{\times}$ & $n^4\bar{n}^3\bar{b}$ & 6.92$^{\times}$ & $s^4\bar{s}^3\bar{c}$ & 7.00$^{\times}$ & $s^4\bar{s}^3\bar{b}$ & 6.95$^{\times}$ & $c^4\bar{c}^3\bar{b}$ & 5.49 & $n^2c^2\bar{n}^2\bar{b}^2$ & 6.27$^{\times}$ \\ \hline
        7-quark
        & $c^5\bar{c}^2$  & 5.41 & $b^5\bar{b}^2$ & 3.97 & & & & & & & $nc^3b\bar{b}^2$ & 5.24 \\
        & $nc^4\bar{c}^2$ & 5.65$^{\ast}$ & $nb^4\bar{b}^2$ & 4.65 & $sc^4\bar{c}^2$ & 5.67$^{\ast}$ & $sb^4\bar{b}^2$ & 4.70 & $cb^4\bar{b}^2$ & 4.28 & $nc^4\bar{b}^2$ & 5.39 \\
        & $n^2c^3\bar{c}^2$ & 5.88$^{\ast}$ & $n^2b^3\bar{b}^2$ & 5.19 & $s^2c^3\bar{c}^2$ & 5.92$^{\ast}$ & $s^2b^3\bar{b}^2$ & 5.26 & $c^2b^3\bar{b}^2$ & 4.54 & $n^2c^2b\bar{b}^2$ & 5.51 \\
        & $n^3c^2\bar{c}^2$ & 6.11$^{\times}$ & $n^3b^2\bar{b}^2$ & 5.64$^{\ast}$ & $s^3c^2\bar{c}^2$ & 6.14$^{\times}$ & $s^3b^2\bar{b}^2$ & 5.71$^{\ast}$ & $c^3b^2\bar{b}^2$ & 4.76 & $n^2c^3\bar{b}^2$ & 5.65$^{\ast}$ \\
        & $n^4c\bar{c}^2$ & 6.32$^{\times}$ & $n^4b\bar{b}^2$ & 6.02$^{\times}$ & $s^4c\bar{c}^2$ & 6.36$^{\times}$ & $s^4b\bar{b}^2$ & 6.08$^{\times}$ & $c^4b\bar{b}^2$ & 4.95 & $n^3cb\bar{b}^2$ & 5.77$^{\ast}$ \\
        & $n^5\bar{c}^2$ & 6.51$^{\times}$ & $n^5\bar{b}^2$ & 6.34$^{\times}$ & $s^5\bar{c}^2$ & 6.55$^{\times}$ & $s^5\bar{b}^2$ & 6.39$^{\times}$ & $c^5\bar{b}^2$ & 5.12 & $n^3cc\bar{b}^2$ & 5.89$^{\ast}$ \\
        & $n^5\bar{n}\bar{c}$ & 6.70$^{\times}$ & $n^5\bar{n}\bar{b}$ & 6.63$^{\times}$ & $s^5\bar{s}\bar{c}$ & 6.74$^{\times}$ & $s^5\bar{s}\bar{b}$ & 6.67$^{\times}$ & $c^5\bar{c}\bar{b}$ & 5.27 & $n^4c\bar{b}^2$ & 6.12$^{\times}$ \\ \hline
        6-quark
        & $c^3\bar{c}^3$ & 5.19 & $b^3\bar{b}^3$ & 3.74 & & & & & & & $nc^2\bar{b}^3$ & 4.96 \\
        & $nc^2\bar{c}^3$ & 5.45 & $nb^2\bar{b}^3$ & 4.49 & $sc^2\bar{c}^3$ & 5.47 & $sb^2\bar{b}^3$ & 4.55 & $cb^2\bar{b}^3$ & 4.10 & $c^3\bar{n}\bar{b}^2$ & 5.14 \\
        & $n^2c\bar{c}^3$ & 5.70$^{\ast}$ & $n^2b\bar{b}^3$ & 5.08 & $s^2c\bar{c}^3$ & 5.74$^{\ast}$ & $s^2b\bar{b}^3$ & 5.16 & $c^2b\bar{b}^3$ & 4.39 & $n^2c\bar{b}^3$ & 5.27 \\
        & $n^3\bar{c}^3$ & 5.94$^{\ast}$ & $n^3\bar{b}^3$ & 5.56 & $s^3\bar{c}^3$ & 5.99$^{\ast}$ & $s^3\bar{b}^3$ & 5.63$^{\ast}$ & $c^3\bar{b}^3$ & 4.63 & $c^3\bar{n}^2\bar{b}$ & 5.57 \\
        & $n^3\bar{n}\bar{c}^2$ & 6.17$^{\times}$ & $n^3\bar{n}\bar{b}^2$ & 5.95$^{\ast}$ & $s^3\bar{s}\bar{c}^2$ & 6.22$^{\times}$ & $s^3\bar{s}\bar{b}^2$ & 6.02$^{\times}$ & $c^3\bar{c}\bar{b}^2$ & 4.84 & $n^2c\bar{n}\bar{b}^2$ & 5.70$^{\ast}$ \\
        & $n^3\bar{n}^2\bar{c}$ & 6.38$^{\times}$ & $n^3\bar{n}^2\bar{b}$ & 6.29$^{\times}$ & $s^3\bar{s}^2\bar{c}$ & 6.43$^{\times}$ & $s^3\bar{s}^2\bar{b}$ & 6.34$^{\times}$ & $c^3\bar{c}^2\bar{b}$ & 5.02 & $nc^2\bar{n}^2\bar{b}$ & 5.82$^{\ast}$ \\
    \hline\hline
    \end{tabular}
\end{table*}

Finally, one can search for the multiquark states within the limit derived from bag model.
The predicted bag radii $R_0$ listed in Table \ref{tab:prospect} for possible 6-, 7-, and 8-quark configurations with baryon number $B=0,1$. 
As shown, the values of $R_0$ are highly sensitive to the number of light quarks in flavor constituents due to significant relativistic effects.
The presence of strangeness has minimal impact, altering the bag radius by only about 0.01--0.08$\,\rm{GeV}^{-1}$.
Flavor configurations in Table \ref{tab:prospect} are selected based on permutations of heavy-light combinations to illustrate varying trends in $R_0$, 
including fully heavy and cross-flavor systems (e.g., $c^3\bar{b}^3$, $nc^2\bar{b}^3$). Configurations involving different color representations, 
such as $qqqqqq$ for 6-quark states or $qqqqqqq \bar{q}$ for 8-quark states, are also accounted for via equivalence like $R_0(qqqqqq) = R_0(qqq\bar{q}\bar{q}\bar{q})$, 
since $R_0$ is determined by kinetic and confinement energies, 
with chromomagnetic interactions treated as small perturbations compared to the large hadron masses.
For the same reason, spin splittings are neglected in Table \ref{tab:prospect}.

Applying the critical bag radius $R_c$, we classify states marked with symbol $\ast$
as not recommended for study in compact pictures, due to the positive confinement energy $E_{\rm{CON}}$.
Considering the more precise critical string distance $r_c=1.2$--$1.4\,$fm, 
we exclude many flavor configurations marked with $\times$ from quark confinement, such as doubly charmed hexaquarks $n^3\bar{n}\bar{c}^2$
with bag radius 6.17$\,$GeV$^{-1}$(1.22$\,$fm), close to the RMS radius of 1.1--1.6$\,$fm for $n^4c^2$ in Ref.\cite{An:2025rjv}.
Accordingly, searches for most compact multiquark states with 6, 7, and 8 quarks are not recommended,
except $nc^2\bar{c}^3$, $c^3\bar{c}^3$, $c^5\bar{c}^2$, and some bottom partners.
We note that the flavor configurations excluded in Table \ref{tab:prospect} may still be possible to be molecular or resonant states.
In the next section, we delve into the discussion of hexaquarks with OZI-superallowed strong decay modes.

\section{Hexaquarks with strong decays}
\label{sec:hexaquark}

Apart from limitations imposed by quark confinement, the stability of hexaquarks can also be influenced by OZI-superallowed decay modes \cite{jaffe:1976ig}.
Unlike string-breaking and creation of new $q\bar{q}$ pairs in confinement scenarios, this decay mode does not require exceeding the bag radius or string distance.
Instead, it depends on the color-spin wavefunction, which can separate the quarks into isolated color-singlet substructures, regardless of how small the bag radius is.
Accordingly, we continue our discussion of stability for states not excluded from confinement, as listed in Table \ref{tab:prospect}.

In general, color-spin wavefunctions are employed to exclude scattering states and evaluate relative decay widths of OZI-superallowed decay modes 
by examining the components of color-singlet substructures \cite{zhang:2023teh,Liu:2024mwn,Liu:2025fbe,Liu:2021gva,An:2022fvs,An:2022qpt,Weng:2019ynv,Weng:2022ohh}.
The color-spin wavefunction of a multiquark state can be transformed into representations of meson-meson, meson-baryon, and baryon-baryon couplings for tetraquarks, pentaquarks, and hexaquarks, respectively.
We express the hexaquark wavefunction $\left|\psi\right\rangle$ in terms of color-singlet $\mathbf{1_c}$ and color-octet $\mathbf{8_c}$ components as follows:
\begin{equation}
    \left|\psi\right\rangle = c^{1} \left|q_{1}q_{2}q_{3}\right\rangle^{\mathbf{1}}_{S_1} \left|\bar{q_{4}}\bar{q_{5}}\bar{q_{6}}\right\rangle^{\mathbf{1}}_{S_2} +
    c^{8} \left|q_{1}q_{2}q_{3}\right\rangle^{\mathbf{8}}_{S_3} \left|\bar{q_{4}}\bar{q_{5}}\bar{q_{6}}\right\rangle^{\mathbf{8}}_{S_4} + \dots. 
    \label{equ:component}
\end{equation}
Here, the first term indicates dissociation of state into $S$-wave baryon and antibaryon, with a particular final-state spin configuration $\{S_1,S_2\}$ and wavefunction overlap $c^1$.
The second term represents a compact state strongly bound by gluon exchange between quarks.
Obviously, if the probability $|c^1|^2$ approaches 1, suggesting strong coupling to a scattering state, 
the state must be disregard due to the inappropriate physical picture and broad decay width.

The key question is how to infer a broad decay width from the wavefunction (\ref{equ:component}), which is central to hexaquark stability.
Based on previous studies \cite{Liu:2021gva,An:2022fvs,An:2022qpt,Weng:2019ynv,Weng:2022ohh}, we adopt the formulism for two-body $L$-wave decay widths \cite{gaoc1992}
\begin{equation}
    \Gamma_i = \gamma_i \alpha \frac{k^{2L+1}}{m^{2L}} \cdot |c^1_i|^2,
    \label{equ:partialwidth}
\end{equation}
where $\Gamma_i$ is the partial width for channel $i$, $\gamma_i$ relates to decay dynamics, $\alpha$ is the effective coupling constant,
$m$ is the initial-state mass, $k$ is the momentum of final states in the rest frame, and $c_i^1$ is the overlap $c^1$ for the $\mathbf{1_c}$ component in Eq.(\ref{equ:component}).
For ground-state hexaquark decays, ${(k/m)}^2$ is on the order of $\mathcal{O}(10^{-2})$, so we focus on $S$-wave decays, as higher wave modes are suppressed.
Equation (\ref{equ:partialwidth}) then simplifies to 
\begin{equation}
    \Gamma_i = \gamma_i \alpha k\cdot |c^1_i|^2,
    \label{equ:partialwidthS}
\end{equation}
where $k$ satisfies the conservation equation
\begin{equation}
    m_A = \sqrt{m_B^2+k^2} + \sqrt{m_C^2+k^2},
    \label{equ:momentumk}
\end{equation}
for the process $A\to B+C$, and $\gamma_i$ depends on the spatial wavefunctions of initial and final states.
Currently, relative decay widths can be roughly estimated as the ratio of factors $k\cdot |c_i^1|^2$, since determining $\gamma_i$ and $\alpha$ precisely is challenging \cite{gaoc1992}.
The partial decay width $\Gamma_i$ is sensitive to $|c_i^1|^2$, which can amplify the result, especially for scattering states where $|c_i^1|^2 \approx 1$,
serving as the origin of broad decay widths. To explore the dependence of stability on $|c_i^1|^2$,
we regard the partial width $\Gamma_i$ to be of order $\mathcal{O}(k\cdot |c_i^1|^2)$, due to the same order $\mathcal{O}(1)$ of $\gamma_i$ and $\alpha$ for two-body strong decays, according to Ref. \cite{gaoc1992}.

\subsection{The fully heavy systems}
\label{sec:fullyheavy}

\renewcommand{\tabcolsep}{0.3cm}
\renewcommand{\arraystretch}{1.2}
\begin{table}[!htbp]
    \caption{Predicted mass $M$ (GeV), bag radius $R_{0}$ (GeV$^{-1}$) and factor $k\cdot|c^1_i|^2$ (GeV) for 
        hexaquarks $bbb\bar{b}\bar{b}\bar{b}$, $ccc\bar{b}\bar{b}\bar{b}$ and $ccc\bar{c}\bar{c}\bar{c}$.}
    \label{tab:QQQQQQ}
    \begin{tabular}{cccccc}
        \hline\hline
        System &$J^{P(C)}$ &$R_0$ &$M$ &$|c^1_i|^2$ &$k\cdot|c^1_i|^2$ \\ \hline
        $bbb\bar{b}\bar{b}\bar{b}$
            &${0}^{-+}$ &3.74 &29.803 &72.4$\%$ &2.064 \\
            &           & &29.785 &27.6$\%$ &0.775 \\ \hline
        $ccc\bar{b}\bar{b}\bar{b}$
            &${1}^{-}$  &4.63 &19.904 &18.5$\%$ &0.334 \\
            &${0}^{-}$  & &19.894 &61.2$\%$ &1.093 \\
            &           & &19.866 &38.4$\%$ &0.662 \\ \hline
        $ccc\bar{c}\bar{c}\bar{c}$
            &${0}^{-+}$ &5.19 &9.918 &72.4$\%$ &0.779 \\
            &           & &9.849 &27.6$\%$ &0.250 \\
        \hline\hline
    \end{tabular}
\end{table}

\renewcommand{\tabcolsep}{0.2cm}
\renewcommand{\arraystretch}{1.2}
\begin{table}[!htbp]
    \caption{Predicted mass $M$ (GeV), bag radius $R_{0}$ (4.10 GeV$^{-1}$) and factor $k\cdot|c^1_i|^2$ (GeV) for hexaquarks $bbc\bar{b}\bar{b}\bar{b}$.}
    \label{tab:bbcbbb}
    \begin{tabular}{ccccccc}
        \hline\hline
        \multicolumn{2}{l}{$bbc\bar{b}\bar{b}\bar{b}$} &\multicolumn{2}{c}{$|c^1_i|^2$} &\multicolumn{2}{c}{$k\cdot|c^1_i|^2$} \\
        \cline{3-4} \cline{5-6}
        $J^{P}$ &$M$ &$\Omega^{\ast}_{bbc}\bar{\Omega}_{bbb}$ &$\Omega_{bbc}\bar{\Omega}_{bbb}$ 
        &$\Omega^{\ast}_{bbc}\bar{\Omega}_{bbb}$ &$\Omega_{bbc}\bar{\Omega}_{bbb}$ \\ \hline
            ${2}^{-}$   &26.506 &78.2$\%$ &2.5$\%$ &1.946 &0.063 \\
                        &26.498 &21.1$\%$ &21.1$\%$ &0.520 &0.537 \\
                        &26.483 &0.7$\%$ &76.4$\%$ &0.016 &1.911 \\
            ${1}^{-}$   &26.516 &26.8$\%$ &0.2$\%$ &0.673 &0.006 \\
                        &26.508 &9.1$\%$ &0.1$\%$ &0.226 &0.003 \\
                        &26.501 &50.3$\%$ &2.1$\%$ &1.244 &0.054 \\
                        &26.493 &11.9$\%$ &41.6$\%$ &0.291 &1.053 \\
                        &26.482 &2.0$\%$ &55.9$\%$ &0.049 &1.396 \\
            ${0}^{-}$   &26.512 &69.2$\%$ & &1.730 \\
                        &26.489 &18.2$\%$ & &0.443 \\
                        &26.488 &12.7$\%$ & &0.309 \\
        \hline\hline
    \end{tabular}
\end{table}

\renewcommand{\tabcolsep}{0.2cm}
\renewcommand{\arraystretch}{1.2}
\begin{table}[!htbp]
    \caption{Predicted mass $M$ (GeV), bag radius $R_{0}$ (4.39 GeV$^{-1}$) and factor $k\cdot|c^1_i|^2$ (GeV) for hexaquarks $ccb\bar{b}\bar{b}\bar{b}$.}
    \label{tab:ccbbbb}
    \begin{tabular}{ccccccc}
        \hline\hline
        \multicolumn{2}{l}{$ccb\bar{b}\bar{b}\bar{b}$} &\multicolumn{2}{c}{$|c^1_i|^2$} &\multicolumn{2}{c}{$k\cdot|c^1_i|^2$} \\
        \cline{3-4} \cline{5-6}
        $J^{P}$ &$M$ &$\Omega^{\ast}_{ccb}\bar{\Omega}_{bbb}$ &$\Omega_{ccb}\bar{\Omega}_{bbb}$ 
        &$\Omega^{\ast}_{ccb}\bar{\Omega}_{bbb}$ &$\Omega_{ccb}\bar{\Omega}_{bbb}$ \\ \hline
            ${2}^{-}$   &23.201 &62.8$\%$ &2.2$\%$ &1.358 &0.048 \\
                        &23.194 &36.8$\%$ &8.3$\%$ &0.790 &0.183 \\
                        &23.182 &0.4$\%$ &89.5$\%$ &0.008 &1.938 \\
            ${1}^{-}$   &23.215 &22.1$\%$ &0.7$\%$ &0.485 &0.015 \\
                        &23.202 &7.5$\%$ &8.2$\%$ &0.162 &0.181 \\
                        &23.194 &60.6$\%$ &4.2$\%$ &1.300 &0.092 \\
                        &23.187 &8.5$\%$ &45.1$\%$ &0.181 &0.984 \\
                        &23.174 &1.4$\%$ &41.8$\%$ &0.029 &0.898 \\
            ${0}^{-}$   &23.209 &54.6$\%$ & &1.192 \\
                        &23.194 &17.3$\%$ & &0.371 \\
                        &23.180 &28.1$\%$ & &0.594 \\
        \hline\hline
    \end{tabular}
\end{table}

\renewcommand{\tabcolsep}{0.2cm}
\renewcommand{\arraystretch}{1.2}
\begin{table}[!htbp]
    \caption{Predicted mass $M$ (GeV), bag radius $R_{0}$ (4.84 GeV$^{-1}$) and factor $k\cdot|c^1_i|^2$ (GeV) for hexaquarks $bbc\bar{c}\bar{c}\bar{c}$.}
    \label{tab:bbcccc}
    \begin{tabular}{ccccccc}
        \hline\hline
        \multicolumn{2}{l}{$bbc\bar{c}\bar{c}\bar{c}$} &\multicolumn{2}{c}{$|c^1_i|^2$} &\multicolumn{2}{c}{$k\cdot|c^1_i|^2$} \\
        \cline{3-4} \cline{5-6}
        $J^{P}$ &$M$ &$\Omega^{\ast}_{bbc}\bar{\Omega}_{ccc}$ &$\Omega_{bbc}\bar{\Omega}_{ccc}$ 
        &$\Omega^{\ast}_{bbc}\bar{\Omega}_{ccc}$ &$\Omega_{bbc}\bar{\Omega}_{ccc}$ \\ \hline
            ${2}^{-}$   &16.569 &30.9$\%$ &19.3$\%$ &0.465 &0.303 \\
                        &16.557 &65.3$\%$ &23.5$\%$ &0.962 &0.361 \\
                        &16.536 &3.8$\%$ &57.2$\%$ &0.054 &0.853 \\
            ${1}^{-}$   &16.586 &18.3$\%$ &0.7$\%$ &0.282 &0.012 \\
                        &16.580 &8.2$\%$ &0.5$\%$ &0.126 &0.008 \\
                        &16.563 &21.9$\%$ &21.7$\%$ &0.326 &0.336 \\
                        &16.550 &40.5$\%$ &44.3$\%$ &0.590 &0.675 \\
                        &16.532 &11.0$\%$ &32.8$\%$ &0.155 &0.486 \\
            ${0}^{-}$   &16.575 &65.2$\%$ & &0.988 \\
                        &16.546 &3.9$\%$ & &0.056 \\
                        &16.529 &31.0$\%$ & &0.435 \\
        \hline\hline
    \end{tabular}
\end{table}

\renewcommand{\tabcolsep}{0.2cm}
\renewcommand{\arraystretch}{1.2}
\begin{table}[!htbp]
    \caption{Predicted mass $M$ (GeV), bag radius $R_{0}$ (5.02 GeV$^{-1}$) and factor $k\cdot|c^1_i|^2$ (GeV) for hexaquarks $ccb\bar{c}\bar{c}\bar{c}$.}
    \label{tab:ccbccc}
    \begin{tabular}{ccccccc}
        \hline\hline
        \multicolumn{2}{l}{$ccb\bar{c}\bar{c}\bar{c}$} &\multicolumn{2}{c}{$|c^1_i|^2$} &\multicolumn{2}{c}{$k\cdot|c^1_i|^2$} \\
        \cline{3-4} \cline{5-6}
        $J^{P}$ &$M$ &$\Omega^{\ast}_{ccb}\bar{\Omega}_{ccc}$ &$\Omega_{ccb}\bar{\Omega}_{ccc}$ 
        &$\Omega^{\ast}_{ccb}\bar{\Omega}_{ccc}$ &$\Omega_{ccb}\bar{\Omega}_{ccc}$ \\ \hline
            ${2}^{-}$   &13.244 &23.0$\%$ &10.5$\%$ &0.296 &0.140 \\
                        &13.229 &73.3$\%$ &14.1$\%$ &0.917 &0.184 \\
                        &13.215 &3.7$\%$ &75.4$\%$ &0.045 &0.957 \\
            ${1}^{-}$   &13.262 &25.7$\%$ &2.0$\%$ &0.342 &0.027 \\
                        &13.251 &4.2$\%$ &14.0$\%$ &0.054 &0.189 \\
                        &13.228 &20.3$\%$ &30.7$\%$ &0.254 &0.400 \\
                        &13.219 &44.9$\%$ &28.2$\%$ &0.550 &0.361 \\
                        &13.192 &4.9$\%$ &25.0$\%$ &0.057 &0.303 \\
            ${0}^{-}$   &13.253 &53.1$\%$ & &0.695 \\
                        &13.233 &19.5$\%$ & &0.247 \\
                        &13.191 &27.4$\%$ & &0.316 \\
        \hline\hline
    \end{tabular}
\end{table}

Upon completing the numerical calculations, we first discuss the mass spectra and decay factors for fully heavy hexaquark systems. 
We focus on higher-symmetry configurations such as $bbb\bar{b}\bar{b}\bar{b}$, $ccc\bar{b}\bar{b}\bar{b}$, and $ccc\bar{c}\bar{c}\bar{c}$, which provide sufficient insight into stability.
Lower-symmetry cases like $bbc\bar{b}\bar{b}\bar{c}$ and $ccb\bar{c}\bar{c}\bar{b}$ lead to various mass splittings across up to eight multiplets (as suggested by Appendix B),
potentially complicating experimental and theoretical analyses. Here, we aim to assess the potential for discovering fully heavy hexaquarks.

The predicted masses of hexaquarks $bbb\bar{b}\bar{b}\bar{b}$, $ccc\bar{b}\bar{b}\bar{b}$, and $ccc\bar{c}\bar{c}\bar{c}$ are presented explicitly in Table \ref{tab:QQQQQQ},
classified in spin-parity $J^{P(C)}$, with scattering states disregarded. 
Returning to our spectral studies in Ref.\cite{zhang:2023hmg}, we deduced two linear relations among fully heavy systems as functions of heavy quark number $N$:
\begin{equation}
    \begin{aligned} 
        \bar{M}_{b}\left(N\right) &=5.1004 N-0.7229, \\
        \bar{M}_{c}\left(N\right) &=1.7079 N-0.3241,
    \end{aligned}
    \label{equ:linearMbar}
\end{equation}
in the unified framework of MIT bag model. Following the relations, the masses of hexaquarks $bbb\bar{b}\bar{b}\bar{b}$ and $ccc\bar{c}\bar{c}\bar{c}$ can be roughly estimated as
29.880$\,$GeV and 9.923$\,$GeV, respectively, compared to 29.803$\,$GeV and 9.918$\,$GeV in this work. The discrepancy arises from the approximately linear behavior of fully heavy spectra
demonstrated in the bag model \cite{zhang:2023hmg}. Therefore, prospects of states beyond hexaquarks can be inferred using average masses:
\begin{equation}
    \begin{aligned} 
        &\bar{M}_{b}\left(7\right) =34.980, \quad \bar{M}_{c}\left(7\right) =11.631, \\
        &\bar{M}_{b}\left(8\right) =40.080, \quad \bar{M}_{c}\left(8\right) =13.339,
    \end{aligned}
    \label{equ:prospectMbar}
\end{equation}
such that the mass of a fully heavy 7-quark state falls in the range 11.631--34.980$\,$GeV, while that of an 8-quark state is 13.339--40.080$\,$GeV.
We also examine potentially compact hexaquarks $bbc\bar{b}\bar{b}\bar{b}$, $ccb\bar{b}\bar{b}\bar{b}$, $bbc\bar{c}\bar{c}\bar{c}$, and $ccb\bar{c}\bar{c}\bar{c}$,
as shown in Tables \ref{tab:bbcbbb}, \ref{tab:ccbbbb}, \ref{tab:bbcccc} and \ref{tab:ccbccc}, respectively.
The mass gaps are found to be narrow due to the suppression caused by heavy quarks.

Given the results below the critical bag radius, 
we look forward to the findings of compact fully heavy hexaquarks in OZI-superallowed decay channel.
The fully charm tetraquark $X(6600)$ was reported in di-$J/\psi$ channel 
with broad decay widths of $124^{+32}_{-26}\pm 33\,$MeV or $440^{+230+110}_{-200-240}\,$MeV \cite{CMS:2023owd},
while we evaluate the factor $k\cdot |c_i^1|^2$ to be 114.7$\,$MeV for the $X(6600)$ state according to Ref.\cite{yan:2023lvm}.
Similarly, we perform numerical calculations of factors for fully heavy hexaquarks in their corresponding decay channels, using color-spin wavefunctions.
The masses of triply heavy baryons follow Ref.\cite{zhang:2023hmg}.
As shown in Tables \ref{tab:QQQQQQ}-\ref{tab:ccbccc}, most states may exhibit broad decay widths due to large momentum $k$ in the rest frame and significant probability $|c_i^1|^2$.
We denote states as symbol $H_{}(J^{P(C)},M)$ and expect the discoveries of some particular states such as $H_{bbc\bar{b}\bar{b}\bar{b}}(1^{-},26.508)$, 
$H_{ccb\bar{b}\bar{b}\bar{b}}(1^{-},23.202)$, $H_{bbc\bar{c}\bar{c}\bar{c}}(1^{-},16.580)$, $H_{bbc\bar{c}\bar{c}\bar{c}}(0^{-},16.546)$, $H_{ccb\bar{c}\bar{c}\bar{c}}(1^{-},13.251)$,
$H_{ccb\bar{c}\bar{c}\bar{c}}(0^{-},13.233)$, and $H_{ccc\bar{c}\bar{c}\bar{c}}(0^{-+},9.849)$. States with factors above 1$\,$GeV are not recommended for study due to broad decay widths.

Notably, large values of $|c_i^1|^2$ are typically caused by flavor symmetry (heavy flavor identity) in the wavefunction. Taking $ccc\bar{c}\bar{c}\bar{c}$ as an example, 
there are 72.4$\%$ and 27.6$\%$, both highly significant compared to 3.9$\%$ for the lower-symmetry state $H_{bbc\bar{c}\bar{c}\bar{c}}(0^{-},16.546)$.
Additionally, heavy systems could result in a large momentum $k$, like 1.075$\,$GeV for $H_{ccc\bar{c}\bar{c}\bar{c}}(0^{-+},9.918)$.
The presence of light quarks will suppress $k$ and thus narrow the decay width.

\subsection{The partially heavy systems}
\label{sec:partialheavy}

In this section, we turn to partially heavy systems involving non-strange light quarks.
Replacing with strange quarks minimally affects stability conclusions, as discussed in Table \ref{tab:prospect}, due to the SU(3)$_{f}$ flavor symmetry.
Potentially compact hexaquarks include $bbn\bar{b}\bar{b}\bar{b}$, $ccn\bar{c}\bar{c}\bar{c}$, $nnb\bar{b}\bar{b}\bar{b}$, and $nnn\bar{b}\bar{b}\bar{b}$.
Cross-flavor constituents like $ccn\bar{b}\bar{b}\bar{b}$ or $nnc\bar{b}\bar{b}\bar{b}$ are neglected. Here, we aim to investigate the stability in the presence of light quarks.

\renewcommand{\tabcolsep}{0.25cm}
\renewcommand{\arraystretch}{1.2}
\begin{table}[!htbp]
    \caption{Predicted mass $M$ (GeV), bag radius $R_{0}$ (4.49 GeV$^{-1}$) and factor $k\cdot|c^1_i|^2$ (GeV) for hexaquarks $bbn\bar{b}\bar{b}\bar{b}$.}
    \label{tab:bbnbbb}
    \begin{tabular}{ccccccc}
        \hline\hline
        \multicolumn{2}{l}{$bbn\bar{b}\bar{b}\bar{b}$} &\multicolumn{2}{c}{$|c^1_i|^2$} &\multicolumn{2}{c}{$k\cdot|c^1_i|^2$} \\
        \cline{3-4} \cline{5-6}
        $J^{P}$ &$M$ &$\Xi^{\ast}_{bb}\bar{\Omega}_{bbb}$ &$\Xi_{bb}\bar{\Omega}_{bbb}$ 
        &$\Xi^{\ast}_{bb}\bar{\Omega}_{bbb}$ &$\Xi_{bb}\bar{\Omega}_{bbb}$ \\ \hline
            ${2}^{-}$   &25.420 &78.3$\%$ &1.9$\%$ &1.804 &0.046 \\
                        &25.396 &20.3$\%$ &25.0$\%$ &0.454 &0.592 \\
                        &25.366 &1.4$\%$ &73.1$\%$ &0.031 &1.673 \\
            ${1}^{-}$   &25.435 &20.5$\%$ &0.3$\%$ &0.482 &0.007 \\
                        &25.422 &33.9$\%$ &0.2$\%$ &0.783 &0.005 \\
                        &25.408 &29.0$\%$ &0.2$\%$ &0.658 &0.004 \\
                        &25.386 &13.4$\%$ &35.0$\%$ &0.297 &0.819 \\
                        &25.368 &3.2$\%$ &64.3$\%$ &0.070 &1.475 \\
            ${0}^{-}$   &25.428 &66.9$\%$ & &1.558 \\
                        &25.387 &29.9$\%$ & &0.663 \\
                        &25.370 &3.1$\%$ & &0.068 \\
        \hline\hline
    \end{tabular}
\end{table}

\renewcommand{\tabcolsep}{0.3cm}
\renewcommand{\arraystretch}{1.2}
\begin{table}[!htbp]
    \caption{Predicted mass $M$ (GeV), bag radius $R_{0}$ (5.45 GeV$^{-1}$) and factor $k\cdot|c^1_i|^2$ (GeV) for hexaquarks $ccn\bar{c}\bar{c}\bar{c}$.}
    \label{tab:ccnccc}
    \begin{tabular}{cccccc}
        \hline\hline
        \multicolumn{2}{l}{$ccn\bar{c}\bar{c}\bar{c}$} &\multicolumn{2}{c}{$|c^1_i|^2$} &\multicolumn{2}{c}{$k\cdot|c^1_i|^2$} \\
        \cline{3-4} \cline{5-6}
        $J^{P}$ &$M$ &$\Xi^{\ast}_{cc}\bar{\Omega}_{ccc}$ &$\Xi_{cc}\bar{\Omega}_{ccc}$ 
        &$\Xi^{\ast}_{cc}\bar{\Omega}_{ccc}$ &$\Xi_{cc}\bar{\Omega}_{ccc}$ \\ \hline
            ${2}^{-}$   &8.735 &78.3$\%$ &2.3$\%$ &0.685 &0.025 \\
                        &8.691 &20.9$\%$ &22.1$\%$ &0.159 &0.216 \\
                        &8.615 &0.8$\%$ &75.7$\%$ &0.004 &0.606 \\
            ${1}^{-}$   &8.782 &25.6$\%$ &0.2$\%$ &0.252 &0.003 \\
                        &8.743 &13.9$\%$ &\textless0.1$\%$ &0.124 &\textless0.001 \\
                        &8.709 &46.0$\%$ &1.5$\%$ &0.372 &0.016 \\
                        &8.665 &12.2$\%$ &40.5$\%$ &0.083 &0.374 \\
                        &8.610 &2.3$\%$ &57.6$\%$ &0.011 &0.454 \\
            ${0}^{-}$   &8.760 &68.6$\%$ & &0.641 &\\
                        &8.646 &31.2$\%$ & &0.194 & \\
                        &8.638 &0.2$\%$ & &0.001 & \\
        \hline\hline
    \end{tabular}
\end{table}

\renewcommand{\tabcolsep}{0.1cm}
\renewcommand{\arraystretch}{1.2}
\begin{table}[!htbp]
    \caption{Predicted mass $M$ (GeV), bag radius $R_{0}$ (5.08 GeV$^{-1}$) and factor $k\cdot|c^1_i|^2$ (GeV) for hexaquarks $nnb\bar{b}\bar{b}\bar{b}$.}
    \label{tab:nnbbbb}
    \begin{tabular}{ccccccccc}
        \hline\hline
        \multicolumn{3}{l}{$nnb\bar{b}\bar{b}\bar{b}$} &\multicolumn{3}{c}{$|c^1_i|^2$} &\multicolumn{3}{c}{$k\cdot|c^1_i|^2$} \\
        \cline{4-6} \cline{7-9}
        $I$ &$J^{P}$ &$M$ &$\Sigma^{\ast}_{b}\bar{\Omega}_{bbb}$ &$\Sigma_{b}\bar{\Omega}_{bbb}$ &$\Lambda_{b}\bar{\Omega}_{bbb}$ 
        &$\Sigma^{\ast}_{b}\bar{\Omega}_{bbb}$ &$\Sigma_{b}\bar{\Omega}_{bbb}$ &$\Lambda_{b}\bar{\Omega}_{bbb}$ \\ \hline
        1
            &${2}^{-}$  &20.892 &61.1$\%$ &1.1$\%$ & &1.172 &0.021 & \\
            &           &20.872 &38.4$\%$ &5.7$\%$ & &0.720 &0.109 & \\
            &${1}^{-}$  &21.098 &0.7$\%$ &0.3$\%$ & &0.016 &0.008 \\
            &           &20.894 &41.3$\%$ &3.2$\%$ & &0.795 &0.062 & \\
            &           &20.878 &50.2$\%$ &7.0$\%$ & &0.948 &0.136 & \\
            &           &20.858 &7.5$\%$ &20.3$\%$ & &0.139 &0.383 & \\
            &           &20.842 &0.3$\%$ &69.0$\%$ & &0.005 &1.273 & \\
            &${0}^{-}$  &21.092 &1.3$\%$ & & &0.030 \\
        0
            &${2}^{-}$  &20.989 & & &0.3$\%$ & & &0.007 \\
            &${1}^{-}$  &21.009 & & &\textless0.1$\%$ & & &\textless0.001 \\
            &           &20.968 & & &0.3$\%$ & & &0.008 \\
            &           &20.665 & & &79.5$\%$ & & &1.481 \\
            &           &20.662 & & &20.2$\%$ & & &0.375 \\
        \hline\hline
    \end{tabular}
\end{table}

\renewcommand{\tabcolsep}{0.25cm}
\renewcommand{\arraystretch}{1.2}
\begin{table}[!htbp]
    \caption{Predicted mass $M$ (GeV), bag radius $R_{0}$ (5.56 GeV$^{-1}$) and factor $k\cdot|c^1_i|^2$ (GeV) for hexaquarks $nnn\bar{b}\bar{b}\bar{b}$.}
    \label{tab:nnnbbb}
    \begin{tabular}{ccccccc}
        \hline\hline
        \multicolumn{3}{l}{$nnn\bar{b}\bar{b}\bar{b}$} &\multicolumn{2}{c}{$|c^1_i|^2$} &\multicolumn{2}{c}{$k\cdot|c^1_i|^2$} \\
        \cline{4-5} \cline{6-7}
        $I$ &$J^{P}$ &$M$ &$\Delta\bar{\Omega}_{bbb}$ &$N\bar{\Omega}_{bbb}$ &$\Delta\bar{\Omega}_{bbb}$ &$N\bar{\Omega}_{bbb}$ \\ \hline
        $3/2$
            &${1}^{-}$  &16.571 &0.4$\%$ & &0.006 \\
            &${0}^{-}$  &16.522 &1.1$\%$ & &0.015 \\
        $1/2$
            &${2}^{-}$  &16.406 & &0.1$\%$ & &0.001 \\
            &${1}^{-}$  &16.413 & &\textless0.1$\%$ & &\textless0.001 \\
            &           &16.322 & &0.1$\%$ & &0.002 \\
        \hline\hline
    \end{tabular}
\end{table}

Numerical results are listed in Tables \ref{tab:bbnbbb}, \ref{tab:ccnccc}, \ref{tab:nnbbbb}, and \ref{tab:nnnbbb}, 
for $bbn\bar{b}\bar{b}\bar{b}$, $ccn\bar{c}\bar{c}\bar{c}$, $nnb\bar{b}\bar{b}\bar{b}$, and $nnn\bar{b}\bar{b}\bar{b}$, respectively.
Focusing on mass spectra, we find that the mass ranges widen with light quarks compared to the narrow gaps in fully heavy systems.
For instance, the mass range of $bbn\bar{b}\bar{b}\bar{b}$ is 67$\,$MeV, larger than 18$\,$MeV for $bbb\bar{b}\bar{b}\bar{b}$, 
while for $nnb\bar{b}\bar{b}\bar{b}$ it is 256$\,$MeV up to 436$\,$MeV owing to the separation of isospin symmetry.
The range for $ccn\bar{c}\bar{c}\bar{c}$ is 172$\,$MeV compared to 69$\,$MeV for $ccc\bar{c}\bar{c}\bar{c}$.
This arises from suppression of the coupling parameter (\ref{equ:Cij}) by heavy quarks, where the magnetic moment $\bar{\mu}_i$ typically scales as $1/m_i$.
We look forward to the early findings of hexaquarks $ccn\bar{c}\bar{c}\bar{c}$ at lower energies around $8\,$GeV; then, $nnn\bar{b}\bar{b}\bar{b}$ could be discovered in LHCb.

As mentioned, factors $k\cdot |c_i^1|^2$ can be attributed to both phase space and flavor symmetry, characterized by momentum $k$ and probability $|c_i^1|^2$. 
More discoveries of heavy hexaquarks are expected from Tables \ref{tab:bbnbbb}, \ref{tab:ccnccc}, \ref{tab:nnbbbb}, and \ref{tab:nnnbbb},
with light quarks suppressing phase space and reducing the flavor symmetry. 
Notably, $|c_i^1|^2$ can be as small as 3.1$\%$ for $H_{bbn\bar{b}\bar{b}\bar{b}}(0^{-},25.370)$ with a factor of 0.068$\,$GeV, suggesting a possible narrow decay width. 
For the $ccn\bar{c}\bar{c}\bar{c}$ system, the lowest-mass state $H_{ccn\bar{c}\bar{c}\bar{c}}(0^{-},8.638)$ has 0.2$\%$, yielding a tiny factor of 0.001$\,$GeV.
As the number of light quarks increases, more states exhibit tiny factors,
such as $H_{nnb\bar{b}\bar{b}\bar{b}}(01^{-},21.009)$, $H_{nnb\bar{b}\bar{b}\bar{b}}(01^{-},20.968)$, $H_{nnb\bar{b}\bar{b}\bar{b}}(11^{-},21.098)$, and $H_{nnb\bar{b}\bar{b}\bar{b}}(10^{-},21.092)$.
Special attention is given to the $nnn\bar{b}\bar{b}\bar{b}$ system, where all mass splittings indicate possible narrow decay widths, suggesting compact states in experiments.

The connection between $|c_i^1|^2$ and flavor symmetry can be traced to the interaction matrix. 
Consider the subspace $\big(\frac{1}{2}(\phi_{2}\chi_{20}-\phi_{3}\chi_{18}-\phi_{4}\chi_{14}+\phi_{5}\chi_{12}),\phi_{6}\chi_{4}\big)$ as an example, 
involving systems $(bbb\bar{b}\bar{b}\bar{b},0^{-+})$, $(ccc\bar{b}\bar{b}\bar{b},0^{-})$ and $((nnn)^{I=3/2}\bar{b}\bar{b}\bar{b},0^{-})$ in Table \ref{tab:wvs},
with matrices in analytical and numerical forms (in MeV):
\begin{itemize}
    \item $bbb\bar{b}\bar{b}\bar{b}$, $0^{-+}$
    \begin{align}
    E_{\rm{CMI}} &= 
    \begin{pmatrix}
        0 & -16C_{bb} \\
        -16C_{bb} & 16C_{bb}
    \end{pmatrix}
    =
    \begin{pmatrix}
        0 & -7.9 \\
        -7.9 & 7.9
    \end{pmatrix}, \\
    E_{\rm{BD}} &= 
    \begin{pmatrix}
        6B_{bb} & 0 \\
        0 & 6B_{bb}
    \end{pmatrix}
    =
    \begin{pmatrix}
        -767.3 & 0 \\
        0 & -767.3
    \end{pmatrix}.
    \end{align}
    \item $ccc\bar{b}\bar{b}\bar{b}$, $0^{-}$
    \begin{align}
    E_{\rm{CMI}} &= 
    \begin{pmatrix}
        10C_{bb}-20C_{bc}+10C_{cc} & -16C_{bc} \\
        -16C_{bc} & 8C_{bb}+8C_{cc}
    \end{pmatrix} \nonumber \\
    &=
    \begin{pmatrix}
        9.5 & -13.6 \\
        -13.6 & 21.2
    \end{pmatrix}, \\
    E_{\rm{BD}} &= 
    \begin{pmatrix}
        \frac{3}{4}(B_{bb}+6B_{bc}+B_{cc}) & 0 \\
        0 & 3B_{bb}+3B_{cc}
    \end{pmatrix} \nonumber \\
    &=
    \begin{pmatrix}
        -609.8 & 0 \\
        0 & -615.2
    \end{pmatrix}.
    \end{align}
    \item $(nnn)^{I=3/2}\bar{b}\bar{b}\bar{b}$, $0^{-}$
    \begin{align}
    E_{\rm{CMI}} &= 
    \begin{pmatrix}
        10C_{bb}-20C_{bn}+10C_{nn} & -16C_{bn} \\
        -16C_{bn} & 8C_{bb}+8C_{nn}
    \end{pmatrix} \nonumber \\
    &=
    \begin{pmatrix}
        154.2 & -30.7 \\
        -30.7 & 154.1
    \end{pmatrix}, \\
    E_{\rm{BD}} &= 
    \begin{pmatrix}
        \frac{3}{4}B_{bb} & 0 \\
        0 & 3B_{bb}
    \end{pmatrix}
    =
    \begin{pmatrix}
        -95.9 & 0 \\
        0 & -383.6
    \end{pmatrix}.
    \end{align}
\end{itemize}
For the fully heavy system $bbb\bar{b}\bar{b}\bar{b}$ with the highest flavor symmetry, the diagonal elements of $E_{\rm{CMI}}+E_{\rm{BD}}$ differ by only 7.9$\,$MeV, 
leading to strong mixing and significant probability $|c_i^1|^2$ for the color-singlet basis $\phi_6\chi_4$.
Replacing $bbb$ with charm introduces weaker binding energies $B_{bc}$ and $B_{cc}$, reducing the energy difference to 6.3$\,$MeV and increasing $|c_i^1|^2$.
Finally, without the binding energies between light and heavy quarks, 
the 287.8$\,$MeV difference for the $nnn\bar{b}\bar{b}\bar{b}$ system implies negligible mixing between $1_c \otimes 1_c$ and $8_c \otimes 8_c$ bases,
dividing eigenvectors into a scattering state ($|c_i^1|^2=98.9\%$) and a compact state ($|c_i^1|^2=1.1\%$). Apparently, the values of $|c_i^1|^2$ are dominated by the binding energy, 
reflecting heavy flavor identity and short-range volume occupation, and serving as the possible source of broad decay widths for heavy hadrons.

\subsection{The non-linear bag radius}
\label{sec:radius}

\begin{figure*}[t]
    \centering
    \subfigure[$\ $ $N_c$ in fully bottom systems\label{fig:bRNc1}]{\includegraphics[width=0.45\textwidth]{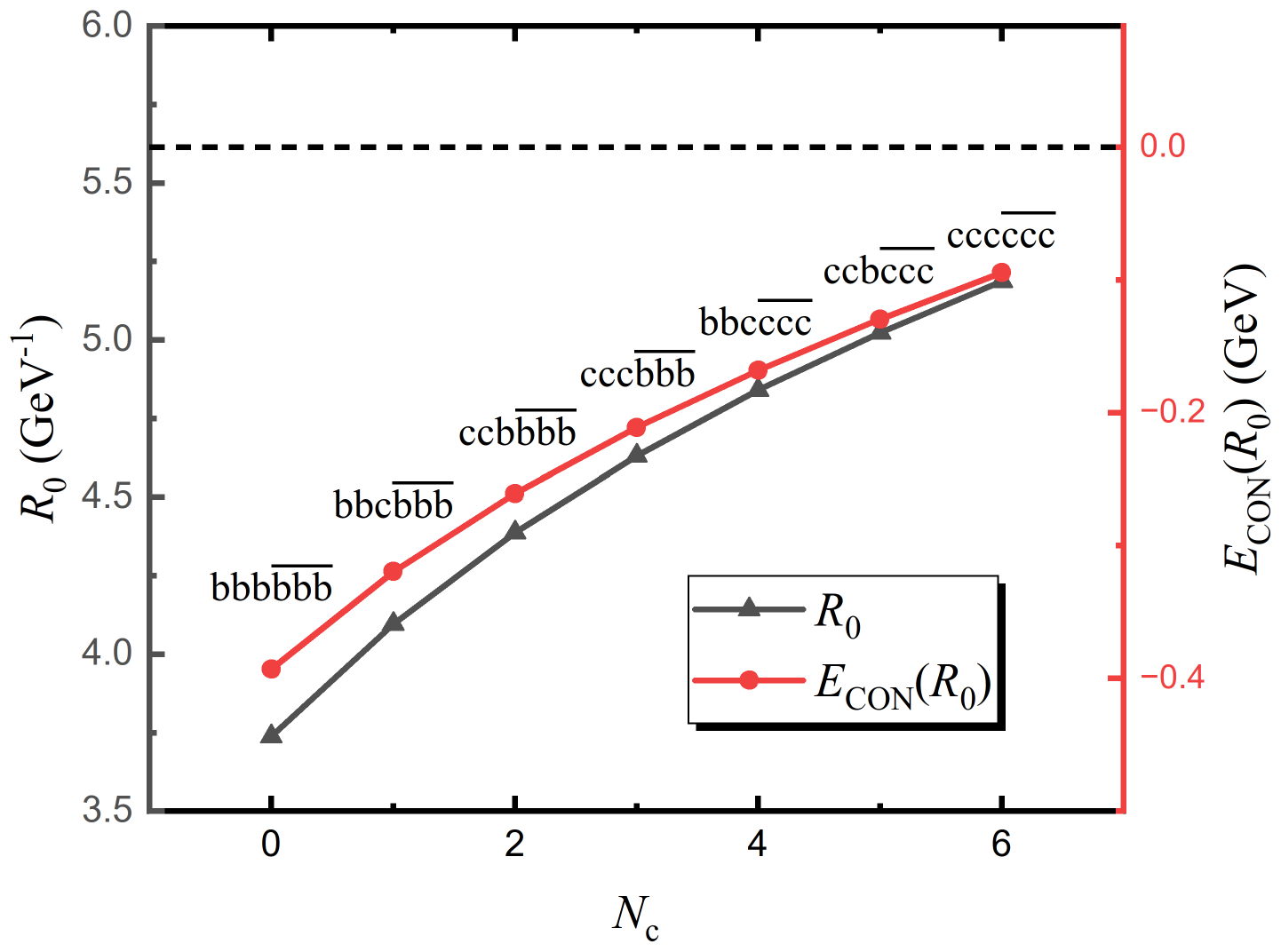}}
    \subfigure[$\ $ $N_s$ in fully non-strange light systems\label{fig:bRNc2}]{\includegraphics[width=0.45\textwidth]{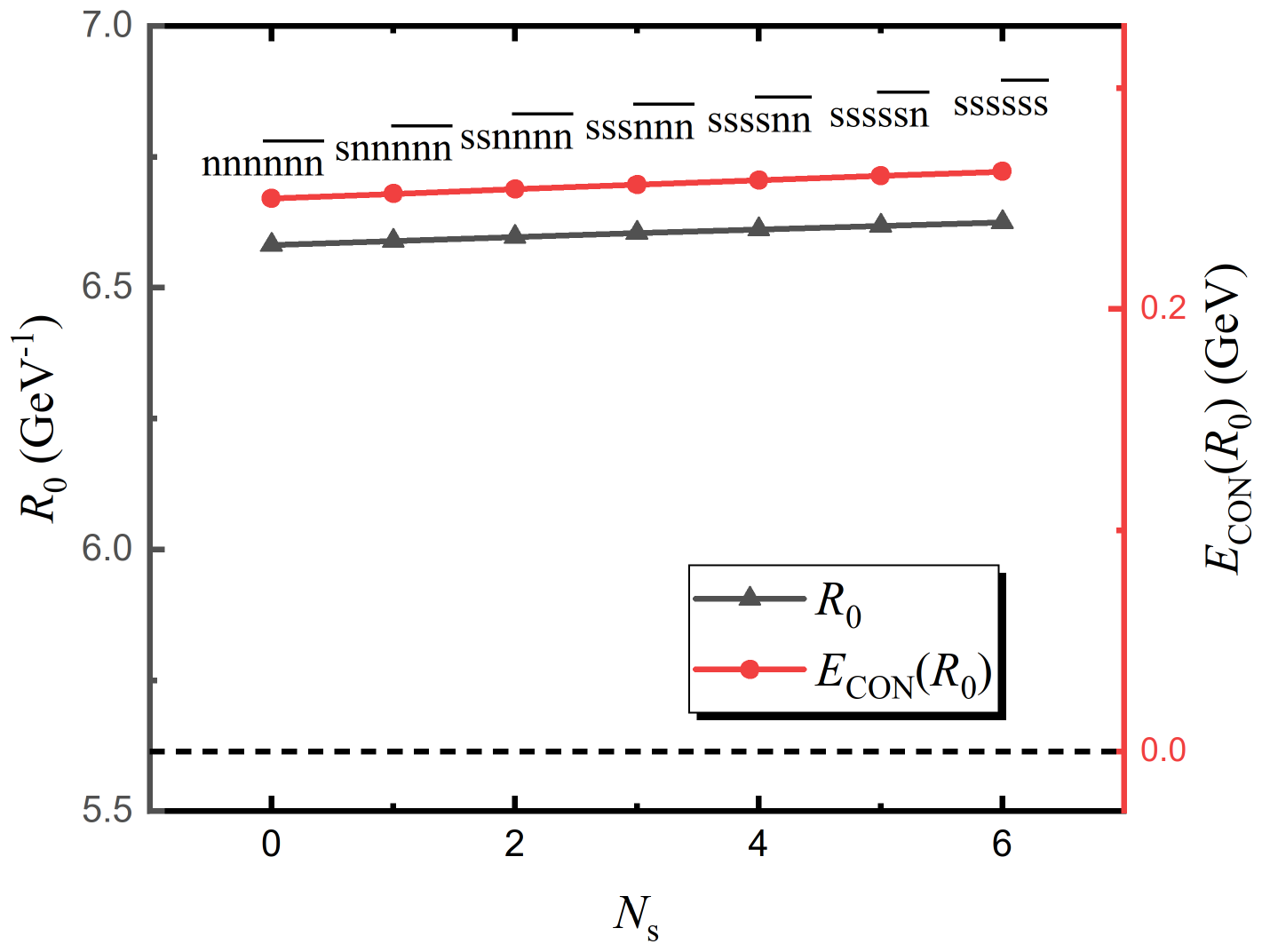}}
    \par
    \subfigure[$\ $ $N_n$ in fully bottom systems\label{fig:bRNc3}]{\includegraphics[width=0.45\textwidth]{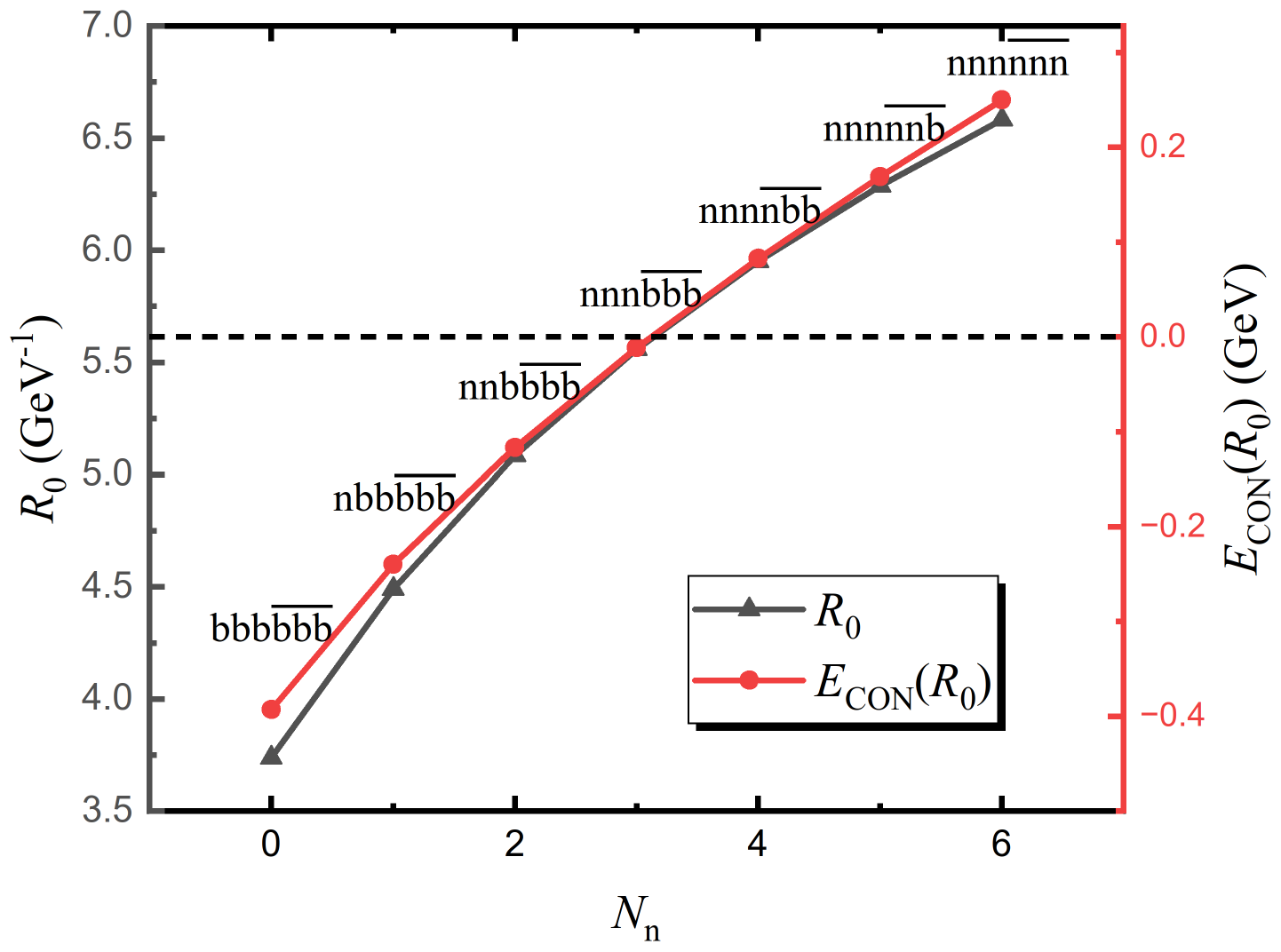}}
    \subfigure[$\ $ $N_n$ in fully charm systems\label{fig:bRNc4}]{\includegraphics[width=0.45\textwidth]{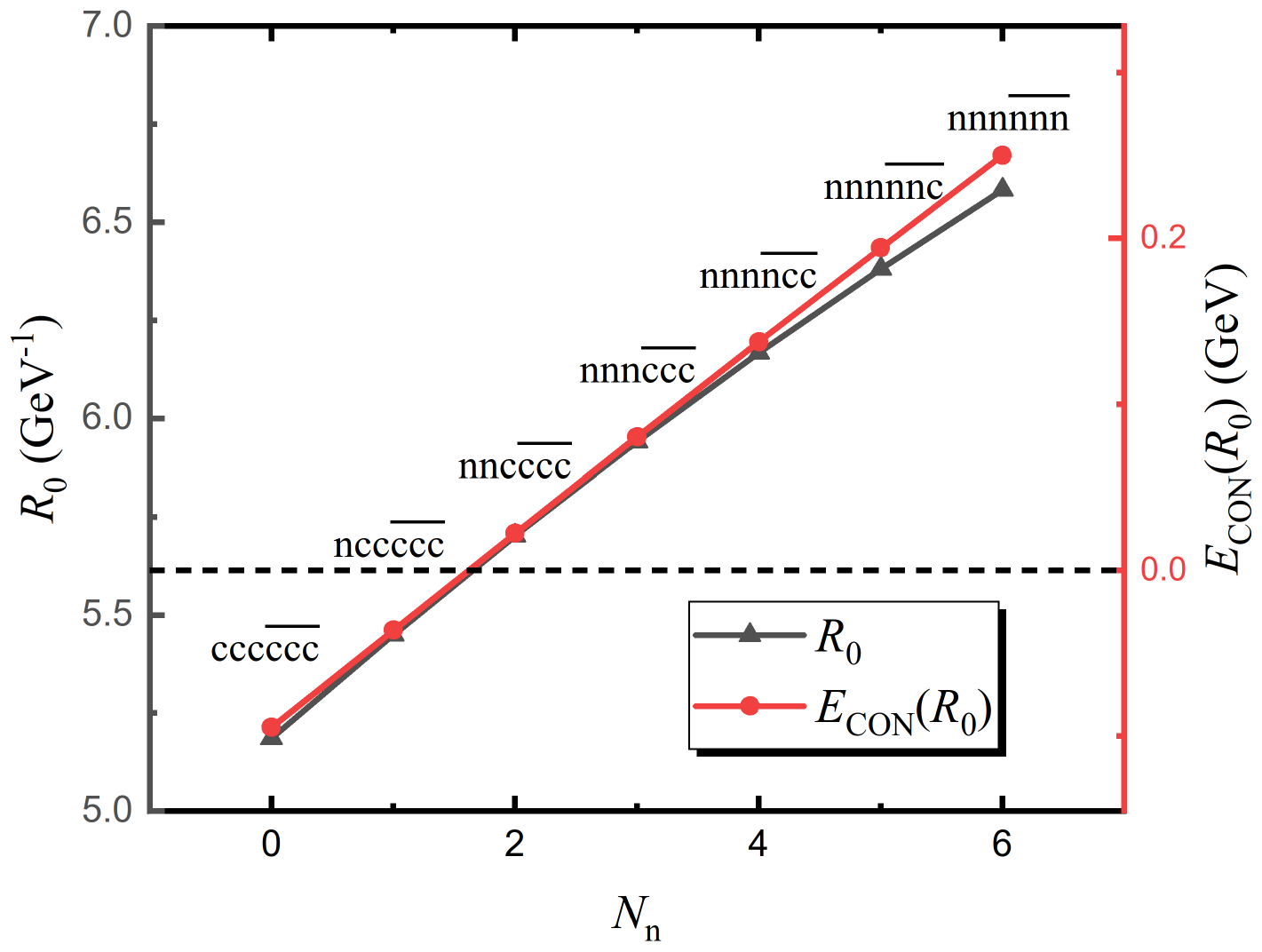}}
    \caption{The plots of bag radii $R_0$ of hexaquark family labeled with gray triangles, and their confinement energy $E_{\rm{CON}}(R_0)$ denoted by red sphere dots,
        arising upon number $N_c$ of charm ($N_s$ of strange, $N_n$ of non-strange light) quarks in flavor configurations. 
        The critical bag radius $R_c=5.61\,\rm{GeV}^{-1}$ aligns with zero $E_{\rm{CON}}$ by a short-dashed line.}
    \label{fig:bRNc}
\end{figure*}

In the final section, we discuss the bag radii obtained above. 
As seen in Tables \ref{tab:prospect}-\ref{tab:ccbccc}, the bag radius varies minimally without light quarks.
It can be as small as 3.74$\,\rm{GeV}^{-1}$ for $bbb\bar{b}\bar{b}\bar{b}$ and up to 5.19$\,\rm{GeV}^{-1}$ for $ccc\bar{c}\bar{c}\bar{c}$, holding within the critical bag radius $R_c$.
Whereas, the bag radius is strongly sensitive to the number of light quarks in flavor configurations, representing a broad range of values.
We plot the corresponding radii $R_0$ with their confinement energy $E_{\rm{CON}}(R_{0})$ in Fig.(\ref{fig:bRNc}), for fully heavy, fully light, and partially heavy systems,
where the curves illustrate the non-linear relation between quark constituents and the energy scale characterized by $R_0$.

This phenomenon aligns with two-body strong interactions, where flavor constituents exhibit a non-linear relation with energy scale
denoted by momentum transfer $Q^2$, reduced mass $\mu_{ij}=m_i m_j/(m_i+m_j)$ or Bohr radius ($\sim 1/r_0$). 
For the simplest two-body strong interaction, heavy mesons $b\bar{b}$, $b\bar{c}$ and $c\bar{c}$ are classified by $N_c=0,1,2$, mimicking Fig.(\ref{fig:bRNc1}),
where the inverse reduced mass $1/\mu_{ij}$ exhibits a concave function upon $N_c$, similar to the radius $R_0$.
The concavity enhances for $b-n$ coupling in Fig.(\ref{fig:bRNc3}), reduces for $c-n$ coupling in Fig.(\ref{fig:bRNc4}), and vanishes for $s-n$ coupling in Fig.(\ref{fig:bRNc2}).
This can be attributed to flavor symmetry breaking. Additionally, the concavity of confinement energy $E_{\rm{CON}}(R_{0})$ suggests a possible origin of 
mass inequality from flavor permutations \cite{Bertlmann:1979zs,Weingarten:1983uj,Witten:1983ut,Nussinov:1983hb,Lieb:1985aw,Martin:1986da,Karliner:2019vhw}.

As noted for the mean field ansatz, the bag model averages and absorbs long-range confinement (vortex) potential within each quark pair
into the overall bag confinement energy, solving short-range perturbative (gluon) interactions independently.
This procedure can be attributed to the challenges in many-body effects and separated energy scales (IR and UV) in QCD, providing general and qualitative conclusions.
Given the similar phenomenon between two-body strong interactions and many-body mean field ansatz in Fig.(\ref{fig:bRNc}), 
we affirm the stability conclusions in this work.

\section{Summary and conclusions}
\label{sec:summary}

In this work, we investigate the prospects for compact hexaquarks and more complex multiquark states, such as 7- and 8-quark configurations,
by imposing a confinement scale within the MIT bag model. This study addresses the challenge of multiquark stability under quark confinement, 
highlighting the underlying limitations of flavor constituents in compact states. 
Using color-spin wavefunctions constructed via Young tableaux,
we compute wavefunction overlaps to gain insights into stability against OZI-superallowed decays. 
The systems examined include fully heavy hexaquarks like $b^3\bar{b}^3$ and $c^3\bar{c}^3$, 
as well as partially heavy hexaquarks such as $n^3\bar{b}^3$ and $n^3\bar{c}^3$. 
Our objective is to assess the possibility of discovering these compact hexaquarks.

Applying the MIT bag model and its associated confinement picture, we decompose the mass equation into kinetic energies, perturbative interactions and confinement energy $E_{\rm{CON}}$. 
The critical bag radius $R_c=5.61\,$GeV$^{-1}$ is derived by satisfying the condition $E_{\rm{CON}}<0$ at zero temperature and zero baryon density, 
which aligns with lattice QCD data on the string-breaking distance of 1.2--1.4$\,$fm. 
The connection between mean field ansatz and two-body strong interactions is thoroughly discussed.
This approach concludes that most heavy-light multiquark states violate the confinement scale, as their bag radii $R_0$ exceed $R_c$, resulting in positive $E_{\rm{CON}}$.
The presence of relativistically light quarks compromises stability under confinement, especially for charm systems such as $n^3\bar{c}^3$ and $n^3\bar{n}\bar{c}^2$. 
In contrast, fully heavy systems and certain bottom partners exhibit smaller radii, supporting potentially compact scenarios for multiquark states.

The stability against OZI-superallowed decays is characterized by wavefunction overlaps, particularly the probability $|c_i^1|^2$ of collapsing into the $1_c \otimes 1_c$ component. 
Broad decay widths in fully heavy systems for OZI-superallowed modes may arise from large wavefunction overlaps due to heavy flavor symmetry and substantial binding energies,
reflecting the short-range volume occupation by heavy quarks. Conversely, partially heavy systems may exhibit narrow decay widths, 
as light quarks introduce a lack of binding energy and reduced phase space. As the number of light quarks decreases,
heavy-light hexaquarks exhibit smaller $|c_i^1|^2$ values, such as 3.1$\%$ for $bbn\bar{b}\bar{b}\bar{b}$, 0.3$\%$ for $nnb\bar{b}\bar{b}\bar{b}$ and 0.1$\%$ for $nnn\bar{b}\bar{b}\bar{b}$,
suggesting possible narrow decay widths for OZI-superallowed decay modes.

Overall, our study indicates that light quarks in flavor constituents lead to violations of the confinement scale, resulting in unstable configurations, 
while enabling possible narrow decay widths for partially heavy systems in OZI-superallowed modes. Fully heavy hexaquarks are expected at high energies above 9.8$\,$GeV from $ccc\bar{c}\bar{c}\bar{c}$,
with broad decay widths. Partially heavy hexaquarks of interest include $ccn\bar{c}\bar{c}\bar{c}$ at lower energies around 8.7$\,$GeV, 
as well as $nnb\bar{b}\bar{b}\bar{b}$ and $nnn\bar{b}\bar{b}\bar{b}$, which may be discovered as compact states with narrow decay widths.
We hope that our conclusions regarding the confinement scale and decay properties will guide future searches for hexaquarks.

\medskip \textbf{ACKNOWLEDGMENTS}

W.-X. Zhang thanks Ming-Zhu Liu for the valuable discussions on mean field ansatz and the understanding of two-body strong interactions.
D. J. is supported by the National Natural Science Foundation of China under Grant No. 12165017.
\iffalse
This work is supported by the National Natural Science Foundation of China under Grant Nos. 12335001, 12247155, 12247101, and 12405097, 
National Key Research and Development Program of China under Contract No. 2020YFA0406400, the ‘111 Center’ under Grant No. B20063,
the Natural Science Foundation of Gansu Province (No. 22JR5RA389), the fundamental Research Funds for the Central Universities, and the project for top-notch innovative talents of Gansu province.
\fi

\medskip
\section*{Appendix A: Color and Spin Wavefunctions}
\label{apd:basis}
\setcounter{equation}{0}
\renewcommand{\theequation}{A\arabic{equation}}

In the framework of MIT bag model, color-spin wavefunctions are required to average the Hamiltonian and obtain the interaction matrices,
as in the chromomagnetic interaction approach. For a hexaquark with baryon number $B=0$, the color configuration could be either baryon-antibaryon coupling 
$(3_c \otimes 3_c \otimes 3_c) \otimes (\bar{3}_c \otimes \bar{3}_c \otimes \bar{3}_c)$ or trimeson coupling $(3_c \otimes \bar{3}_c) \otimes (3_c \otimes \bar{3}_c) \otimes (3_c \otimes \bar{3}_c)$.
To study hexaquark stability against flavor symmetry in this work, we apply the first color configuration and decompose it into direct sums:
\begin{equation}
    \begin{aligned}
        &(3_c \otimes 3_c \otimes 3_c) \otimes (\bar{3}_c \otimes \bar{3}_c \otimes \bar{3}_c) \\
        =& \big[(\bar{3}_c \oplus 6_c) \otimes 3_c\big] \otimes \big[(3_c \oplus \bar{6}_c) \otimes \bar{3}_c\big] \\
        =& (1_c \oplus 8_c \oplus 8_c \oplus 10_c) \otimes (1_c \oplus 8_c \oplus 8_c \oplus \bar{10}_c). \label{equ:colors}
    \end{aligned}
\end{equation}
From Eq.(\ref{equ:colors}), we obtain color-singlet, color-decuplet, and two octets from $\bar{3}_c \otimes 3_c$ and $6_c \otimes 3_c$, 
offering six color bases for the hexaquark:
\begin{align}
    \phi_{1} &= | [(12)^{6}3]^{10} [(\bar{4}\bar{5})^{\bar{6}}\bar{6}]^{\bar{10}} \rangle, \nonumber \\
    \phi_{2} &= | [(12)^{6}3]^{8} [(\bar{4}\bar{5})^{\bar{6}}\bar{6}]^{8} \rangle, \nonumber \\
    \phi_{3} &= | [(12)^{6}3]^{8} [(\bar{4}\bar{5})^{3}\bar{6}]^{8} \rangle, \nonumber \\
    \phi_{4} &= | [(12)^{\bar{3}}3]^{8} [(\bar{4}\bar{5})^{\bar{6}}\bar{6}]^{8} \rangle, \nonumber \\
    \phi_{5} &= | [(12)^{\bar{3}}3]^{8} [(\bar{4}\bar{5})^{3}\bar{6}]^{8} \rangle, \nonumber \\
    \phi_{6} &= | [(12)^{\bar{3}}3]^{1} [(\bar{4}\bar{5})^{3}\bar{6}]^{1} \rangle. \label{equ:colorwv}
\end{align}
Apart from the first five states that couple two colored substructures, particular attention is given to $\phi_6$ which informs stability research.
The spin bases can be established directly by enumerating all possible combinations: 
\begin{align}
    \chi_{1} &= | [(12)_{1}3]_{3/2} [(\bar{4}\bar{5})_{1}\bar{6}]_{3/2} \rangle_{3}, \nonumber \\
    \chi_{2} &= | [(12)_{1}3]_{3/2} [(\bar{4}\bar{5})_{1}\bar{6}]_{3/2} \rangle_{2}, \nonumber \\
    \chi_{3} &= | [(12)_{1}3]_{3/2} [(\bar{4}\bar{5})_{1}\bar{6}]_{3/2} \rangle_{1}, \nonumber \\
    \chi_{4} &= | [(12)_{1}3]_{3/2} [(\bar{4}\bar{5})_{1}\bar{6}]_{3/2} \rangle_{0}, \nonumber \\
    \chi_{5} &= | [(12)_{1}3]_{3/2} [(\bar{4}\bar{5})_{1}\bar{6}]_{1/2} \rangle_{2}, \nonumber \\
    \chi_{6} &= | [(12)_{1}3]_{3/2} [(\bar{4}\bar{5})_{1}\bar{6}]_{1/2} \rangle_{1}, \nonumber \\
    \chi_{7} &= | [(12)_{1}3]_{3/2} [(\bar{4}\bar{5})_{0}\bar{6}]_{1/2} \rangle_{2}, \nonumber \\
    \chi_{8} &= | [(12)_{1}3]_{3/2} [(\bar{4}\bar{5})_{0}\bar{6}]_{1/2} \rangle_{1}, \nonumber \\
    \chi_{9} &= | [(12)_{1}3]_{1/2} [(\bar{4}\bar{5})_{1}\bar{6}]_{3/2} \rangle_{2}, \nonumber \\
    \chi_{10} &= | [(12)_{1}3]_{1/2} [(\bar{4}\bar{5})_{1}\bar{6}]_{3/2} \rangle_{1}, \nonumber \\
    \chi_{11} &= | [(12)_{1}3]_{1/2} [(\bar{4}\bar{5})_{1}\bar{6}]_{1/2} \rangle_{1}, \nonumber \\
    \chi_{12} &= | [(12)_{1}3]_{1/2} [(\bar{4}\bar{5})_{1}\bar{6}]_{1/2} \rangle_{0}, \nonumber \\
    \chi_{13} &= | [(12)_{1}3]_{1/2} [(\bar{4}\bar{5})_{0}\bar{6}]_{1/2} \rangle_{1}, \nonumber \\
    \chi_{14} &= | [(12)_{1}3]_{1/2} [(\bar{4}\bar{5})_{0}\bar{6}]_{1/2} \rangle_{0}, \nonumber \\
    \chi_{15} &= | [(12)_{0}3]_{1/2} [(\bar{4}\bar{5})_{1}\bar{6}]_{3/2} \rangle_{2}, \nonumber \\
    \chi_{16} &= | [(12)_{0}3]_{1/2} [(\bar{4}\bar{5})_{1}\bar{6}]_{3/2} \rangle_{1}, \nonumber \\
    \chi_{17} &= | [(12)_{0}3]_{1/2} [(\bar{4}\bar{5})_{1}\bar{6}]_{1/2} \rangle_{1}, \nonumber \\
    \chi_{18} &= | [(12)_{0}3]_{1/2} [(\bar{4}\bar{5})_{1}\bar{6}]_{1/2} \rangle_{0}, \nonumber \\
    \chi_{19} &= | [(12)_{0}3]_{1/2} [(\bar{4}\bar{5})_{0}\bar{6}]_{1/2} \rangle_{1}, \nonumber \\
    \chi_{20} &= | [(12)_{0}3]_{1/2} [(\bar{4}\bar{5})_{0}\bar{6}]_{1/2} \rangle_{0}. \label{equ:spinwv}
\end{align}
The subscripts in Eq.(\ref{equ:spinwv}) indicate the spin numbers for two-fermion coupling, three-fermion coupling, and the hexaquark state, respectively.

The hadronic wavefunction can be separated into spatial, flavor, color, and spin components, where the spatial part holds symmetric for the ground state.
The flavor configuration constrains the color-spin wavefunctions due to the Pauli principle.
For the hexaquarks addressed in this work, various structures are possible, including isoscalar, isovector, flavor-octet and decuplet systems.
Here, we present all possible flavor-color-spin wavefunctions in Young tableaux for each case.

SU(3)$_f$ $10_f$ \cite{Wu:2017weo}: $(nnn)^{I=3/2}$, $ccc$, $bbb$, $\bar{c}\bar{c}\bar{c}$, $\bar{b}\bar{b}\bar{b}$

\renewcommand{\tabcolsep}{0.1cm}
\renewcommand{\arraystretch}{1}
\begin{align}
    \frac{1}{\sqrt{2}}
    \begin{tabular}{|c|c|c|}
        \cline{1-3}
        1 & 2 & 3 \\
        \cline{1-3}
    \end{tabular}_{f}
    \otimes(
    \begin{tabular}{|c|c|}
        \cline{1-2}
        1 & 2   \\
        \cline{1-2}
        \multicolumn{1}{|c|}{3} \\
        \cline{1-1}
    \end{tabular}_{c}
    \otimes
    \begin{tabular}{|c|c|}
        \cline{1-2}
        1 & 3    \\
        \cline{1-2}
        \multicolumn{1}{|c|}{2} \\
        \cline{1-1}
    \end{tabular}_{s}
    -
    \begin{tabular}{|c|c|}
        \cline{1-2}
        1 & 3   \\
        \cline{1-2}
        \multicolumn{1}{|c|}{2} \\
        \cline{1-1}
    \end{tabular}_{c}
    \otimes
    \begin{tabular}{|c|c|}
        \cline{1-2}
        1 & 2    \\
        \cline{1-2}
        \multicolumn{1}{|c|}{3} \\
        \cline{1-1}
    \end{tabular}_{s}
    ), \label{equ:young1}
\end{align}

\renewcommand{\tabcolsep}{0.1cm}
\renewcommand{\arraystretch}{1}
\begin{align}
    \begin{tabular}{|c|c|c|}
        \cline{1-3}
        1 & 2 & 3 \\
        \cline{1-3}
    \end{tabular}_{f}
    \otimes
    \begin{tabular}{|c|}
        \cline{1-1}
        1 \\
        \cline{1-1}
        2 \\
        \cline{1-1}
        3 \\
        \cline{1-1}
    \end{tabular}_{c}
    \otimes
    \begin{tabular}{|c|c|c|}
        \cline{1-3}
        1 & 2 & 3 \\
        \cline{1-3}
    \end{tabular}_{s}. \label{equ:young2}
\end{align}

SU(3)$_f$ $8_f$ \cite{Wu:2017weo}: $(nnn)^{I=1/2}$

\renewcommand{\tabcolsep}{0.1cm}
\renewcommand{\arraystretch}{1}
\begin{align}
    \frac{1}{\sqrt{2}}
    (
    \begin{tabular}{|c|c|}
        \cline{1-2}
        1 & 2    \\
        \cline{1-2}
        \multicolumn{1}{|c|}{3} \\
        \cline{1-1}
    \end{tabular}_{f}
    \otimes
    \begin{tabular}{|c|c|}
        \cline{1-2}
        1 & 3   \\
        \cline{1-2}
        \multicolumn{1}{|c|}{2} \\
        \cline{1-1}
    \end{tabular}_{c}
    -
    \begin{tabular}{|c|c|}
        \cline{1-2}
        1 & 3    \\
        \cline{1-2}
        \multicolumn{1}{|c|}{2} \\
        \cline{1-1}
    \end{tabular}_{f}
    \otimes
    \begin{tabular}{|c|c|}
        \cline{1-2}
        1 & 2   \\
        \cline{1-2}
        \multicolumn{1}{|c|}{3} \\
        \cline{1-1}
    \end{tabular}_{c}
    )
    \otimes
    \begin{tabular}{|c|c|c|}
        \cline{1-3}
        1 & 2 & 3 \\
        \cline{1-3}
    \end{tabular}_{s}, \label{equ:young3}
\end{align}

\renewcommand{\tabcolsep}{0.1cm}
\renewcommand{\arraystretch}{1}
\begin{align}
    \frac{1}{2}
    [(
    \begin{tabular}{|c|c|}
        \cline{1-2}
        1 & 2    \\
        \cline{1-2}
        \multicolumn{1}{|c|}{3} \\
        \cline{1-1}
    \end{tabular}_{f}
    \otimes
    \begin{tabular}{|c|c|}
        \cline{1-2}
        1 & 3   \\
        \cline{1-2}
        \multicolumn{1}{|c|}{2} \\
        \cline{1-1}
    \end{tabular}_{s}
    +
    \begin{tabular}{|c|c|}
        \cline{1-2}
        1 & 3    \\
        \cline{1-2}
        \multicolumn{1}{|c|}{2} \\
        \cline{1-1}
    \end{tabular}_{f}
    \otimes
    \begin{tabular}{|c|c|}
        \cline{1-2}
        1 & 2   \\
        \cline{1-2}
        \multicolumn{1}{|c|}{3} \\
        \cline{1-1}
    \end{tabular}_{s}
    )
    \otimes
    \begin{tabular}{|c|c|}
        \cline{1-2}
        1 & 2   \\
        \cline{1-2}
        \multicolumn{1}{|c|}{3} \\
        \cline{1-1}
    \end{tabular}_{c} \nonumber\\
    +
    (
    \begin{tabular}{|c|c|}
        \cline{1-2}
        1 & 2    \\
        \cline{1-2}
        \multicolumn{1}{|c|}{3} \\
        \cline{1-1}
    \end{tabular}_{f}
    \otimes
    \begin{tabular}{|c|c|}
        \cline{1-2}
        1 & 2   \\
        \cline{1-2}
        \multicolumn{1}{|c|}{3} \\
        \cline{1-1}
    \end{tabular}_{s}
    -
    \begin{tabular}{|c|c|}
        \cline{1-2}
        1 & 3    \\
        \cline{1-2}
        \multicolumn{1}{|c|}{2} \\
        \cline{1-1}
    \end{tabular}_{f}
    \otimes
    \begin{tabular}{|c|c|}
        \cline{1-2}
        1 & 3   \\
        \cline{1-2}
        \multicolumn{1}{|c|}{2} \\
        \cline{1-1}
    \end{tabular}_{s}
    )
    \otimes
    \begin{tabular}{|c|c|}
        \cline{1-2}
        1 & 3   \\
        \cline{1-2}
        \multicolumn{1}{|c|}{2} \\
        \cline{1-1}
    \end{tabular}_{c}
    ], \label{equ:young4}
\end{align}

\renewcommand{\tabcolsep}{0.1cm}
\renewcommand{\arraystretch}{1}
\begin{align}
    \frac{1}{\sqrt{2}}
    (
    \begin{tabular}{|c|c|}
        \cline{1-2}
        1 & 2    \\
        \cline{1-2}
        \multicolumn{1}{|c|}{3} \\
        \cline{1-1}
    \end{tabular}_{f}
    \otimes
    \begin{tabular}{|c|}
        \cline{1-1}
        1 \\
        \cline{1-1}
        2 \\
        \cline{1-1}
        3 \\
        \cline{1-1}
    \end{tabular}_{c}
    \otimes
    \begin{tabular}{|c|c|}
        \cline{1-2}
        1 & 2   \\
        \cline{1-2}
        \multicolumn{1}{|c|}{3} \\
        \cline{1-1}
    \end{tabular}_{s}
    +
    \begin{tabular}{|c|c|}
        \cline{1-2}
        1 & 3    \\
        \cline{1-2}
        \multicolumn{1}{|c|}{2} \\
        \cline{1-1}
    \end{tabular}_{f}
    \otimes
    \begin{tabular}{|c|}
        \cline{1-1}
        1 \\
        \cline{1-1}
        2 \\
        \cline{1-1}
        3 \\
        \cline{1-1}
    \end{tabular}_{c}
    \otimes
    \begin{tabular}{|c|c|}
        \cline{1-2}
        1 & 3   \\
        \cline{1-2}
        \multicolumn{1}{|c|}{2} \\
        \cline{1-1}
    \end{tabular}_{s}
    ). \label{equ:young5}
\end{align}

SU(2)$_f$ $3_f$: $(nn)^{I=1}$, $cc$, $bb$, $\bar{c}\bar{c}$, $\bar{b}\bar{b}$

\renewcommand{\tabcolsep}{0.1cm}
\renewcommand{\arraystretch}{1}
\begin{align}
    \begin{tabular}{|c|c|}
        \cline{1-2}
        1 & 2 \\
        \cline{1-2}
    \end{tabular}_{f}
    \otimes
    \begin{tabular}{|c|}
        \cline{1-1}
        1 \\
        \cline{1-1}
        2 \\
        \cline{1-1}
    \end{tabular}_{c}
    \otimes
    \begin{tabular}{|c|c|}
        \cline{1-2}
        1 & 2 \\
        \cline{1-2}
    \end{tabular}_{s}
    , \quad
    \begin{tabular}{|c|c|}
        \cline{1-2}
        1 & 2 \\
        \cline{1-2}
    \end{tabular}_{f}
    \otimes
    \begin{tabular}{|c|c|}
        \cline{1-2}
        1 & 2 \\
        \cline{1-2}
    \end{tabular}_{c}
    \otimes
    \begin{tabular}{|c|}
        \cline{1-1}
        1 \\
        \cline{1-1}
        2 \\
        \cline{1-1}
    \end{tabular}_{s}. \label{equ:young6}
\end{align}

SU(2)$_f$ $1_f$: $(nn)^{I=0}$

\renewcommand{\tabcolsep}{0.1cm}
\renewcommand{\arraystretch}{1}
\begin{align}
    \begin{tabular}{|c|}
        \cline{1-1}
        1 \\
        \cline{1-1}
        2 \\
        \cline{1-1}
    \end{tabular}_{f}
    \otimes
    \begin{tabular}{|c|}
        \cline{1-1}
        1 \\
        \cline{1-1}
        2 \\
        \cline{1-1}
    \end{tabular}_{c}
    \otimes
    \begin{tabular}{|c|}
        \cline{1-1}
        1 \\
        \cline{1-1}
        2 \\
        \cline{1-1}
    \end{tabular}_{s}
    , \quad
    \begin{tabular}{|c|}
        \cline{1-1}
        1 \\
        \cline{1-1}
        2 \\
        \cline{1-1}
    \end{tabular}_{f}
    \otimes
    \begin{tabular}{|c|c|}
        \cline{1-2}
        1 & 2 \\
        \cline{1-2}
    \end{tabular}_{c}
    \otimes
    \begin{tabular}{|c|c|}
        \cline{1-2}
        1 & 2 \\
        \cline{1-2}
    \end{tabular}_{s}. \label{equ:young7}
\end{align}

Among these expressions, the wavefunctions remain asymmetric under exchange of any quark pair $1\leftrightarrow 2$, $1\leftrightarrow 3$, or $2\leftrightarrow 3$, 
satisfying the Pauli principle. Thus, we construct the possible color-spin 
wavefunctions according to Young tableaux (\ref{equ:young1})--(\ref{equ:young7}), using bases (\ref{equ:colorwv}) and (\ref{equ:spinwv}).
The flavor configurations and corresponding wavefunctions are listed collectively in Table \ref{tab:wvs}, classified by spin-parity $J^{P(C)}$.
We also detail their derivation step by step in the next section.

\renewcommand{\tabcolsep}{0.3cm} \renewcommand{\arraystretch}{1.5}
\begin{table*}[!htb]
    \caption{Color-spin wavefunctions of fully heavy hexaquarks $QQQ\bar{Q}\bar{Q}\bar{Q}$, $QQQ^{\prime}\bar{Q}\bar{Q}\bar{Q}$, $Q^{\prime}Q^{\prime}Q\bar{Q}\bar{Q}\bar{Q}$,
        and $Q^{\prime}Q^{\prime}Q^{\prime}\bar{Q}\bar{Q}\bar{Q}$ ($Q,Q^{\prime}=c,b$), and partially heavy hexaquarks $QQn\bar{Q}\bar{Q}\bar{Q}$, $nnQ\bar{Q}\bar{Q}\bar{Q}$, and $nnn\bar{Q}\bar{Q}\bar{Q}$, 
        classified into spin-parity $J^{P(C)}$.}
    \label{tab:wvs}
    \begin{tabular}{lll}
    \hline\hline
        System &$J^{P(C)}$ &Color-spin wavefunctions \\ \hline
        $QQQ\bar{Q}\bar{Q}\bar{Q}$
        &   $0^{-+}$ & $\frac{1}{2}(\phi_{2}\chi_{20}-\phi_{3}\chi_{18}-\phi_{4}\chi_{14}+\phi_{5}\chi_{12})$, $\phi_{6}\chi_{4}$ \\
        $Q^{\prime}Q^{\prime}Q^{\prime}\bar{Q}\bar{Q}\bar{Q}$, $(nnn)^{I=3/2}\bar{Q}\bar{Q}\bar{Q}$
        &   $1^{-}$ & $\frac{1}{2}(\phi_{2}\chi_{19}-\phi_{3}\chi_{17}-\phi_{4}\chi_{13}+\phi_{5}\chi_{11})$, $\phi_{6}\chi_{3}$ \\
        &   $0^{-}$ & $\frac{1}{2}(\phi_{2}\chi_{20}-\phi_{3}\chi_{18}-\phi_{4}\chi_{14}+\phi_{5}\chi_{12})$, $\phi_{6}\chi_{4}$ \\
        $(nnn)^{I=1/2}\bar{Q}\bar{Q}\bar{Q}$
        &   $2^{-}$ & $\frac{1}{2}(\phi_{3}\chi_{5}-\phi_{2}\chi_{7}-\phi_{5}\chi_{5}+\phi_{4}\chi_{7})$, $\frac{1}{\sqrt{2}}(\phi_{6}\chi_{9}+\phi_{6}\chi_{15})$ \\
        &   $1^{-}$ & $\frac{1}{2}(\phi_{3}\chi_{6}-\phi_{2}\chi_{8}-\phi_{5}\chi_{6}+\phi_{4}\chi_{8})$, 
                $\frac{1}{2\sqrt{2}}(\phi_{2}\chi_{13}+\phi_{2}\chi_{19}-\phi_{3}\chi_{11}-\phi_{3}\chi_{17}$ \\
        &   &   $+\phi_{4}\chi_{13}-\phi_{4}\chi_{19}-\phi_{5}\chi_{11}+\phi_{5}\chi_{17})$, 
                $\frac{1}{\sqrt{2}}(\phi_{6}\chi_{10}+\phi_{6}\chi_{16})$ \\
        $(nn)^{I=1}Q\bar{Q}\bar{Q}\bar{Q}$, $QQn\bar{Q}\bar{Q}\bar{Q}$, $QQQ^{\prime}\bar{Q}\bar{Q}\bar{Q}$, $Q^{\prime}Q^{\prime}Q\bar{Q}\bar{Q}\bar{Q}$
        &   $2^{-}$ & $\frac{1}{\sqrt{2}}(\phi_{4}\chi_{7}-\phi_{5}\chi_{5})$, $\phi_{6}\chi_{2}$, $\phi_{6}\chi_{9}$ \\
        &   $1^{-}$ & $\frac{1}{\sqrt{2}}(\phi_{2}\chi_{19}-\phi_{3}\chi_{17})$, $\frac{1}{\sqrt{2}}(\phi_{4}\chi_{8}-\phi_{5}\chi_{6})$,
            $\frac{1}{\sqrt{2}}(\phi_{4}\chi_{13}-\phi_{5}\chi_{11})$, $\phi_{6}\chi_{3}$, $\phi_{6}\chi_{10}$ \\
        &   $0^{-}$ & $\frac{1}{\sqrt{2}}(\phi_{2}\chi_{20}-\phi_{3}\chi_{18})$, $\frac{1}{\sqrt{2}}(\phi_{4}\chi_{14}-\phi_{5}\chi_{12})$, $\phi_{6}\chi_{4}$ \\
        $(nn)^{I=0}Q\bar{Q}\bar{Q}\bar{Q}$
        &   $2^{-}$ & $\frac{1}{\sqrt{2}}(\phi_{2}\chi_{7}-\phi_{3}\chi_{5})$, $\phi_{6}\chi_{15}$ \\
        &   $1^{-}$ & $\frac{1}{\sqrt{2}}(\phi_{2}\chi_{8}-\phi_{3}\chi_{6})$, $\frac{1}{\sqrt{2}}(\phi_{2}\chi_{13}-\phi_{3}\chi_{11})$, 
            $\frac{1}{\sqrt{2}}(\phi_{4}\chi_{19}-\phi_{5}\chi_{17})$, $\phi_{6}\chi_{16}$ \\
    \hline\hline
    \end{tabular}%
\end{table*}

\section*{Appendix B: Derivation of Color-Spin Wavefunctions}
\label{apd:derivation}
\setcounter{equation}{0}
\renewcommand{\theequation}{B\arabic{equation}}

We first consider the flavor configuration where only the anti-diquark $\{\bar{4}\bar{5}\}$ is symmetric. 
The possible color-spin bases are selected according to the SU(2)$_f$ symmetry (\ref{equ:young6})--(\ref{equ:young7}).

$J^P=3^-$:
\begin{align}
    \phi_{3}\chi_{1}, \phi_{5}\chi_{1}, \phi_{6}\chi_{1}.
\end{align}

$J^P=2^-$:
\begin{align}
    \phi_{1}\chi_{7}, \phi_{2}\chi_{7}, \phi_{3}\chi_{2}, \phi_{3}\chi_{5}, \phi_{3}\chi_{9}, \phi_{3}\chi_{15}, \nonumber \\
    \phi_{4}\chi_{7}, \phi_{5}\chi_{2}, \phi_{5}\chi_{5}, \phi_{5}\chi_{9}, \phi_{5}\chi_{15}, \nonumber \\
    \phi_{6}\chi_{2}, \phi_{6}\chi_{5}, \phi_{6}\chi_{9}, \phi_{6}\chi_{15}.
\end{align}

$J^P=1^-$:
\begin{align}
    \phi_{1}\chi_{8}, \phi_{1}\chi_{13}, \phi_{1}\chi_{19}, \phi_{2}\chi_{8}, \phi_{2}\chi_{13}, \phi_{2}\chi_{19}, \nonumber \\
    \phi_{3}\chi_{3}, \phi_{3}\chi_{6}, \phi_{3}\chi_{10}, \phi_{3}\chi_{11}, \phi_{3}\chi_{16}, \phi_{3}\chi_{17}, \nonumber \\
    \phi_{4}\chi_{8}, \phi_{4}\chi_{13}, \phi_{4}\chi_{19}, \nonumber \\
    \phi_{5}\chi_{3}, \phi_{5}\chi_{6}, \phi_{5}\chi_{10}, \phi_{5}\chi_{11}, \phi_{5}\chi_{16}, \phi_{5}\chi_{17}, \nonumber \\
    \phi_{6}\chi_{3}, \phi_{6}\chi_{6}, \phi_{6}\chi_{10}, \phi_{6}\chi_{11}, \phi_{6}\chi_{16}, \phi_{6}\chi_{17}.
\end{align}

$J^P=0^-$:
\begin{align}
    \phi_{1}\chi_{14}, \phi_{1}\chi_{20}, \phi_{2}\chi_{14}, \phi_{2}\chi_{20}, \phi_{3}\chi_{4}, \phi_{3}\chi_{12}, \phi_{3}\chi_{18}, \nonumber \\
    \phi_{4}\chi_{14}, \phi_{4}\chi_{20}, \phi_{5}\chi_{4}, \phi_{5}\chi_{12}, \phi_{5}\chi_{18}, \nonumber \\
    \phi_{6}\chi_{4}, \phi_{6}\chi_{12}, \phi_{6}\chi_{18}.
\end{align}

Next, set the diquark $[12]$ to be flavor antisymmetric, typically isoscalar, implying the flavor configuration $[12]3\{\bar{4}\bar{5}\}\bar{6}$. 
The color-spin bases reduce to:

$J^P=3^-$:
\begin{align}
    \phi_{3}\chi_{1}.
\end{align}

$J^P=2^-$:
\begin{align}
    \phi_{1}\chi_{7}, \phi_{2}\chi_{7}, \phi_{3}\chi_{2}, \phi_{3}\chi_{5}, \phi_{3}\chi_{9}, \phi_{5}\chi_{15}, \phi_{6}\chi_{15}.
\end{align}

$J^P=1^-$:
\begin{align}
    \phi_{1}\chi_{8}, \phi_{1}\chi_{13}, \phi_{2}\chi_{8}, \phi_{2}\chi_{13}, \phi_{3}\chi_{3}, \phi_{3}\chi_{6}, \phi_{3}\chi_{10}, \phi_{3}\chi_{11}, \nonumber \\
    \phi_{4}\chi_{19}, \phi_{5}\chi_{16}, \phi_{5}\chi_{17}, \phi_{6}\chi_{16}, \phi_{6}\chi_{17}.
\end{align}

$J^P=0^-$:
\begin{align}
    \phi_{1}\chi_{14}, \phi_{2}\chi_{14}, \phi_{3}\chi_{4}, \phi_{3}\chi_{12}, \phi_{4}\chi_{20}, \phi_{5}\chi_{18}, \phi_{6}\chi_{18}.
\end{align}

Conversely, if the diquark $\{12\}$ is symmetric in flavor wavefunction, the bases for $\{12\}3\{\bar{4}\bar{5}\}\bar{6}$ are:

$J^P=3^-$:
\begin{align}
    \phi_{5}\chi_{1}, \phi_{6}\chi_{1}. \label{equ:cs1245_3}
\end{align}

$J^P=2^-$:
\begin{align}
    \phi_{3}\chi_{15}, \phi_{4}\chi_{7}, \phi_{5}\chi_{2}, \phi_{5}\chi_{5}, \phi_{5}\chi_{9}, \nonumber \\
    \phi_{6}\chi_{2}, \phi_{6}\chi_{5}, \phi_{6}\chi_{9}.
\end{align}

$J^P=1^-$:
\begin{align}
    \phi_{1}\chi_{19}, \phi_{2}\chi_{19}, \phi_{3}\chi_{16}, \phi_{3}\chi_{17}, \phi_{4}\chi_{8}, \phi_{4}\chi_{13}, \nonumber \\
    \phi_{5}\chi_{3}, \phi_{5}\chi_{6}, \phi_{5}\chi_{10}, \phi_{5}\chi_{11}, \nonumber \\
    \phi_{6}\chi_{3}, \phi_{6}\chi_{6}, \phi_{6}\chi_{10}, \phi_{6}\chi_{11}.
\end{align}

$J^P=0^-$:
\begin{align}
    \phi_{1}\chi_{20}, \phi_{2}\chi_{20}, \phi_{3}\chi_{18}, \phi_{4}\chi_{14}, \phi_{5}\chi_{4}, \phi_{5}\chi_{12}, \nonumber \\
    \phi_{6}\chi_{4}, \phi_{6}\chi_{12}. \label{equ:cs1245_0}
\end{align}

Particularly, when charge conjugation symmetry applies to $\{12\}3\{\bar{4}\bar{5}\}\bar{6}$, the color-spin wavefunctions are constrained
from the combinations of bases in Eqs.(\ref{equ:cs1245_3})--(\ref{equ:cs1245_0}).

$J^{PC}=3^{--}$:
\begin{align}
    \phi_{5}\chi_{1}, \phi_{6}\chi_{1}.
\end{align}

$J^{PC}=2^{-+}$:
\begin{align}
    \phi_{5}\chi_{2}, \frac{1}{\sqrt{2}}(\phi_{5}\chi_{5}+\phi_{5}\chi_{9}), \phi_{6}\chi_{2}, \frac{1}{\sqrt{2}}(\phi_{6}\chi_{5}+\phi_{6}\chi_{9}).
\end{align}

$J^{PC}=2^{--}$:
\begin{align}
    \frac{1}{\sqrt{2}}(\phi_{5}\chi_{5}-\phi_{5}\chi_{9}), \frac{1}{\sqrt{2}}(\phi_{6}\chi_{5}-\phi_{6}\chi_{9}).
\end{align}

$J^{PC}=1^{-+}$:
\begin{align}
    \frac{1}{\sqrt{2}}(\phi_{5}\chi_{6}+\phi_{5}\chi_{10}), \frac{1}{\sqrt{2}}(\phi_{6}\chi_{6}+\phi_{6}\chi_{10}).
\end{align}

$J^{PC}=1^{--}$:
\begin{align}
    \phi_{1}\chi_{19}, \phi_{2}\chi_{19}, \phi_{5}\chi_{3}, \frac{1}{\sqrt{2}}(\phi_{5}\chi_{6}-\phi_{5}\chi_{10}), \phi_{5}\chi_{11} \nonumber \\
    \phi_{6}\chi_{3}, \frac{1}{\sqrt{2}}(\phi_{6}\chi_{6}-\phi_{6}\chi_{10}), \phi_{6}\chi_{11}.
\end{align}

$J^{PC}=0^{-+}$:
\begin{align}
    \phi_{1}\chi_{20}, \phi_{2}\chi_{20}, \phi_{5}\chi_{4}, \phi_{5}\chi_{12}, \phi_{6}\chi_{4}, \phi_{6}\chi_{12}.
\end{align}

Now, consider the three-body symmetry governed by SU(3)$_f$ Young tableaux (\ref{equ:young1})--(\ref{equ:young5}). 
In this work, we assume antiquarks to be heavy and identical to ensure high symmetry.
Thus, the color-spin wavefunctions for flavor configuration $[12]3\{\bar{4}\bar{5}\bar{6}\}$ are obtained 
by applying Eqs. (\ref{equ:young1}) and (\ref{equ:young2}) to the bases of $[12]3\{\bar{4}\bar{5}\}\bar{6}$.

$J^P=2^-$:
\begin{align}
    \frac{1}{\sqrt{2}}(\phi_{2}\chi_{7}-\phi_{3}\chi_{5}), \phi_{6}\chi_{15}.
\end{align}

$J^P=1^-$:
\begin{align}
    \frac{1}{\sqrt{2}}(\phi_{2}\chi_{8}-\phi_{3}\chi_{6}), \frac{1}{\sqrt{2}}(\phi_{2}\chi_{13}-\phi_{3}\chi_{11}), \nonumber \\
    \frac{1}{\sqrt{2}}(\phi_{4}\chi_{19}-\phi_{5}\chi_{17}), \phi_{6}\chi_{16}.
\end{align}

The same process applies to $\{12\}3\{\bar{4}\bar{5}\bar{6}\}$.

$J^P=2^-$:
\begin{align}
    \frac{1}{\sqrt{2}}(\phi_{4}\chi_{7}-\phi_{5}\chi_{5}), \phi_{6}\chi_{2}, \phi_{6}\chi_{9}.
\end{align}

$J^P=1^-$:
\begin{align}
    \frac{1}{\sqrt{2}}(\phi_{2}\chi_{19}-\phi_{3}\chi_{17}), \frac{1}{\sqrt{2}}(\phi_{4}\chi_{8}-\phi_{5}\chi_{6}), \nonumber \\
    \frac{1}{\sqrt{2}}(\phi_{4}\chi_{13}-\phi_{5}\chi_{11}), \phi_{6}\chi_{3}, \phi_{6}\chi_{10}.
\end{align}

$J^P=0^-$:
\begin{align}
    \frac{1}{\sqrt{2}}(\phi_{2}\chi_{20}-\phi_{3}\chi_{18}), \frac{1}{\sqrt{2}}(\phi_{4}\chi_{14}-\phi_{5}\chi_{12}), \phi_{6}\chi_{4}.
\end{align}

For flavor-octet $(123)_8$, the wavefunctions of $(123)_8\{\bar{4}\bar{5}\bar{6}\}$ are the direct product of Young tableaux (\ref{equ:young1})--(\ref{equ:young2}) and (\ref{equ:young3})--(\ref{equ:young5}).
We apply them to bases where only $\{\bar{4}\bar{5}\}$ is symmetric, and derive the following equations.

$J^P=2^-$:
\begin{align}
    \frac{1}{2}(\phi_{3}\chi_{5}-\phi_{2}\chi_{7}-\phi_{5}\chi_{5}+\phi_{4}\chi_{7}), \frac{1}{\sqrt{2}}(\phi_{6}\chi_{9}+\phi_{6}\chi_{15}).
\end{align}

$J^P=1^-$:
\begin{align}
    \frac{1}{2}(\phi_{3}\chi_{6}-\phi_{2}\chi_{8}-\phi_{5}\chi_{6}+\phi_{4}\chi_{8}), \nonumber \\
    \frac{1}{2\sqrt{2}}(\phi_{2}\chi_{13}+\phi_{2}\chi_{19}-\phi_{3}\chi_{11}-\phi_{3}\chi_{17} \nonumber \\
    +\phi_{4}\chi_{13}-\phi_{4}\chi_{19}-\phi_{5}\chi_{11}+\phi_{5}\chi_{17}), \nonumber \\
    \frac{1}{\sqrt{2}}(\phi_{6}\chi_{10}+\phi_{6}\chi_{16}).
\end{align}

From the selection of bases for $\{12\}3\{\bar{4}\bar{5}\}\bar{6}$, we obtain the wavefunctions of $\{123\}\{\bar{4}\bar{5}\bar{6}\}$ using Eqs. (\ref{equ:young1}) and (\ref{equ:young2}).

$J^P=1^-$:
\begin{align}
    \frac{1}{2}(\phi_{2}\chi_{19}-\phi_{3}\chi_{17}-\phi_{4}\chi_{13}+\phi_{5}\chi_{11}), \phi_{6}\chi_{3}.
\end{align}

$J^P=0^-$:
\begin{align}
    \frac{1}{2}(\phi_{2}\chi_{20}-\phi_{3}\chi_{18}-\phi_{4}\chi_{14}+\phi_{5}\chi_{12}), \phi_{6}\chi_{4}.
\end{align}

Eventually, the $C$-parity is applied to $\{123\}\{\bar{4}\bar{5}\bar{6}\}$.

$J^{PC}=0^{-+}$:
\begin{align}
    \frac{1}{2}(\phi_{2}\chi_{20}-\phi_{3}\chi_{18}-\phi_{4}\chi_{14}+\phi_{5}\chi_{12}), \phi_{6}\chi_{4}.
\end{align}

Notably, all pure $\phi_6$ states, referring to scattering states, are ignored due to the absence of strong interactions between substructures $123$ and $\bar{4}\bar{5}\bar{6}$.
The basis $\phi_{1}$, coupling the color-decuplet in Eq.(\ref{equ:colorwv}), vanishes after matching flavor symmetry.
This is because if the color and flavor wavefunctions are fully symmetric for substructure $\bar{Q}\bar{Q}\bar{Q}$ within basis $\phi_{1}$, 
the spin part has to be fully asymmetric, which violates the spin-SU(2) algebra.


%\bibliographystyle{apsrev4-1}
%\bibliography{hepref}

\end{document}